\documentclass[pdflatex,sn-mathphys-num]{sn-jnl}

\usepackage{graphicx}%
\usepackage{multirow}%
\usepackage{amsmath,amssymb,amsfonts}%
\usepackage{amsthm}%
\usepackage{mathrsfs}%
\usepackage[title]{appendix}%
\usepackage{xcolor}%
\usepackage{textcomp}%
\usepackage{manyfoot}%
\usepackage{booktabs}%
\usepackage{algorithm}%
\usepackage{algorithmicx}%
\usepackage{algpseudocode}%
\usepackage{listings}%
\usepackage{amssymb}
\usepackage{amsmath}
\usepackage{subcaption}
\usepackage{tikz}
\usepackage{pgfplots}
\usepackage{pgfplotstable}
\usepackage{float}
\usepackage{tikz}
\usepackage{amsmath, amssymb}
\usetikzlibrary{positioning, arrows.meta}
\usepackage{graphicx}
\usepackage{subcaption}

\usepackage{algpseudocode} 
\usepackage{pgfplots}
\pgfplotsset{compat=1.18}
\usetikzlibrary{patterns, decorations.pathmorphing}
\pgfplotsset{compat=1.18}
\usetikzlibrary{arrows.meta, positioning}

\theoremstyle{thmstyleone}%
\theoremstyle{thmstyletwo}%

\theoremstyle{thmstylethree}%

\begin{document}

\title[Article Title]{Stochastic Modeling of Composite Interfaces: Sensitivity to Spatial Correlation and Bayesian Identification from Standard Fracture Tests}


\author*[1,2]{\fnm{Elton Donfack-Siewe}}\email{donfackelton@gmail.com}

\author[2]{\fnm{Sylvain Dubreuil} }\email{sylvain.dubreuil@onera.fr}

\author[3]{\fnm{Christian Fagiano} }\email{christian.fagiano@onera.fr}

\author[2]{\fnm{Jérôme Morio} }\email{jerome.morio@onera.fr}

\author[1]{\fnm{Jean-Philippe Navarro} }\email{jean-philippe.navarro@airbus.com}


\affil*[1]{\orgdiv{} \orgname{Airbus Operation SAS}, \orgaddress{\street{316 route de Bayonne, 31060}, \city{Toulouse}, \country{France}}}

\affil[2]{\orgdiv{DTIS}, \orgname{ONERA, Université de Toulouse, 31000}, \orgaddress{\city{Toulouse},  \country{France}}}

\affil[3]{\orgdiv{DMAS}, \orgname{ONERA, Université Paris-Saclay, 92320}, \orgaddress{\city{Châtillon}, \country{France}}}


\abstract{To enable a numerical handling of uncertainties in composite structures, this work presents a stochastic finite‑element framework aimed at improving the reliability assessment of aerospace composites, with particular attention to stiffener debonding. By representing interface variability between laminate parts with spatially correlated random fields, the method aims at considering scattering effect at a higher scale of simulation and testing. A parametric study carried out on standardized Mode I and Mode II fracture tests reveals that the correlation length is the primary driver of observed variability, while the regularity of the covariance kernel has only a marginal impact. To guarantee industrial relevance, we demonstrate that this key parameter can be extracted from experimental fracture data using an Approximate Bayesian Computation approach. The proposed methodology therefore offers a robust route to high‑fidelity virtual testing and to the predictive management of uncertainties in the design of damage‑tolerant composite airframes.}




\maketitle
\section{Introduction}
\label{sec:introduction}

Composite materials are fundamental to modern aerospace structures due to their exceptional specific mechanical properties~\cite{Yu2016}. However, their deployment is complicated by intrinsic heterogeneity and anisotropy~\cite{campbell2006}, stemming from their multiscale architecture~\cite{Laboulfie2023}. Manufacturing processes—such as vacuum molding or autoclaving—introduce variability in pressure and consolidation~\cite{Grimsley2001}, resulting in interface variability between composite parts. While traditional design methodologies mitigate these uncertainties through safety factors~\cite{CMH17}, the industry's shift toward mass optimization demands a transition from empirical margins to physics-based uncertainty quantification at a higher scale of analysis. This statistical approach directly supports the pillars of Integrated Computational Materials Engineering (ICME) by establishing a digital thread between manufacturing-induced heterogeneities and macro-scale structural integrity.

A major difficulty in aerospace design is the assembly of composite parts, especially the interface between the fuselage skin and the stiffener (stringer). In stiffened panels, the overall structural integrity is often dictated more by the behavior of the adhesive or co‑cured bond line than by the bulk material itself~\cite{Yu2016, Meeks2005}. Therefore, the reliability of the entire assembly hinges on accurately predicting the initiation and propagation of debonding at this specific interface. Present practices usually represent this connection with Cohesive Zone Models (CZM) that assume spatially uniform material parameters—fracture energy $G_c$ and peak strength $\sigma_{max}$~\cite{AbaqusCohesive, Alfano2006}.

Nevertheless, modeling the skin‑stringer joint as a perfectly homogeneous interface contradicts actual manufacturing conditions. Standardized characterization tests used in the industry—especially the Mode I and Mode II procedures—regularly show pronounced spatial scatter in fracture propagation and intricate R‑curve behaviors that constant‑parameter models are unable to capture ~\cite{Pereira2004_ModeI,Pereira2003_ModeII,Kozik2017}.

This study investigates the influence of spatial variability on the interface properties of composite components by proposing a stochastic numerical framework. Unlike conventional deterministic models that assume uniform interface properties, our approach employs spatially correlated random fields to represent manufacturing heterogeneities. The objective is to evaluate how this stochastic representation affects the structural response compared to classical constant-property assumptions. This statistical approach aligns with advanced trends in computational mechanics~\cite{Vinot2023, Stefanou2019,VanBavel2023}. Notably, recent studies have demonstrated how random microstructure morphology significantly influences the macroscopic bending behavior of composite plates, further justifying the need for stochastic frameworks~\cite{Gavallas2025,ostoja1994micromechanics}. However, one essential aspect of this work is the identification of these statistical parameters. We address this by employing Bayesian approach~\cite{Thomas2012}.

The proposed framework is illustrated using experimental datasets derived from standard aerospace characterization protocols. While the specific calibration data used in this work originates from legacy industrial standards, the methodology is inherently versatile. The numerical framework remains valid for any source of experimental data and can be seamlessly adapted to alternative identification methodologies without altering the underlying predictive strategy.  As a prerequisite to the identification process, we first perform a sensitivity analysis to determine which statistical features truly govern the risk of debonding, thereby rationalizing the subsequent calibration strategy.

The paper is organized as follows. Section~\ref{sec:cohesive_model} details the cohesive crack model. Section~\ref{sec:proposed_modeling} presents the stochastic framework using random fields applied to the interface as a means to explicitly treat spatial variability in composite interfaces. This section also provides a qualitative comparison between this random field approach for treating interface uncertainties and the classical approach, which assumes uniform interface properties. Section~\ref{sec:result} analyzes the numerical structural sensitivity to spatial correlation using coupon specimens. Finally, Section~\ref{sec:bayesian} details the Bayesian identification methodology applied to industrial datasets to calibrate this stochastic framework followed by conclusions and future perspectives in Section~\ref{sec:conclusion}.

\section{Cohesive crack model}
\label{sec:cohesive_model}
To simulate interfacial damage and fracture mechanisms—such as delamination in composites or debonding in structural assemblies—\textit{cohesive elements} \cite{Barenblatt1962} provide a computationally efficient and numerical framework. They are particularly advantageous for modeling discontinuities along predefined paths without the need for complex dynamic remeshing.

Consider an interface governed by a \textit{Cohesive Zone Model}. This model relates the traction vector acting on the interface surfaces to the displacement jump across them. The constitutive behavior is defined by a traction–separation law, typically expressed as:
\begin{equation}
\boldsymbol{\sigma} = (1 - D) \, \mathbf{K}_0 \, \boldsymbol{\delta},
\end{equation}
where \( \boldsymbol{\sigma} = (\sigma_n, \sigma_s, \sigma_t)^T \) represents the traction vector (comprising normal, tangential components and assuming \( \sigma_s=\sigma_t\)), \( \boldsymbol{\delta} = (\delta_n, \delta_s, \delta_t)^T \) is the displacement jump vector, and \( \mathbf{K}_0 \) is the initial stiffness penalty matrix of the undamaged interface. The scalar variable \( D \in [0,1] \) represents the damage state, evolving from 0 (intact) to 1 (fully failed).

\subsection{Damage initiation and evolution}

Damage initiation is governed by a failure criterion based on the stress state at the interface. A widely adopted quadratic stress criterion is used:
\begin{equation}
\left( \frac{\langle \sigma_n \rangle}{\sigma_{n,c}} \right)^2
+ \left( \frac{\sigma_s}{\sigma_{s,c}} \right)^2
+ \left( \frac{\sigma_t}{\sigma_{t,c}} \right)^2
= 1,
\end{equation}
where $\sigma_{n,c}$, $\sigma_{s,c}$ and $\sigma_{t,c}$ are the critical cohesive strengths in the normal and shear directions. 
The Macaulay bracket $\langle \cdot \rangle$ is defined as:
\begin{equation}
\langle x \rangle = \max(x,0),
\end{equation}
ensuring that compressive normal stresses do not contribute to damage due to crack-closure effects.

\begin{figure}[H]
\centering
\begin{tikzpicture}
  \begin{axis}[
    axis lines=left,
    xlabel={$ \delta_n\ [\text{mm}] $},
    ylabel={$ \sigma_n\ [\text{MPa}] $},
    xtick=\empty,
    ytick=\empty,
    xmin=0, xmax=0.06,
    ymin=0, ymax=70,
    width=10cm, height=6cm,
    clip=false,
  ]

  \def\sigmaMax{60}
  \def\deltaZero{0.01}
  \def\deltaC{0.05}

  \addplot[domain=0:\deltaZero, samples=100, thick] {(\sigmaMax/\deltaZero)*x};
  \addplot[domain=\deltaZero:\deltaC, samples=100, thick] {\sigmaMax*(1 - (x - \deltaZero)/(\deltaC - \deltaZero))};
  \addplot[domain=\deltaC:0.06, samples=2, thick] {0};

  \addplot [
    domain=0:\deltaC,
    samples=200,
    draw=none,
    fill=red,
    fill opacity=0.2,
    pattern=north east lines,
    pattern color=red]
  {x <= \deltaZero ? (\sigmaMax/\deltaZero)*x : \sigmaMax*(1 - (x - \deltaZero)/(\deltaC - \deltaZero))} \closedcycle;

  \addplot[dashed, gray] coordinates {(\deltaZero, 0) (\deltaZero, \sigmaMax)};
  \addplot[dashed, black] coordinates {(\deltaC, 0) (\deltaC, \sigmaMax)};

  \addplot[only marks, mark=*] coordinates {(\deltaZero, \sigmaMax)};
  \node at (axis cs:\deltaZero, \sigmaMax + 5) [anchor=south] {$(\delta_0, \sigma_{n,c})$};

  \node at (axis cs:\deltaZero, -5) [anchor=north] {$\delta_0$};
  \node at (axis cs:\deltaC, -5) [anchor=north] {$\delta_c$};

  \node at (axis cs:{(1.5*\deltaZero + \deltaC)/2}, \sigmaMax/3) [text=red] {$G_{Ic}$};

  \addplot[domain=0:\deltaZero, samples=2, gray, dotted, thick] {(\sigmaMax/\deltaZero)*x};
  \node at (axis cs:\deltaZero/2, \sigmaMax/2 + 5) [anchor=south, text=gray] {$E_n$};
  
  \end{axis}
\end{tikzpicture}
\caption{Bilinear cohesive law describing the Mode I traction $\sigma_n$ versus displacement jump $\delta_n$. The shaded area represents the critical fracture energy $G_{Ic}$. The peak traction $\sigma_{n,c}$ occurs at displacement $\delta_0$, and complete failure occurs at $\delta_c$.}
\label{fig:cohesive_law}
\end{figure}
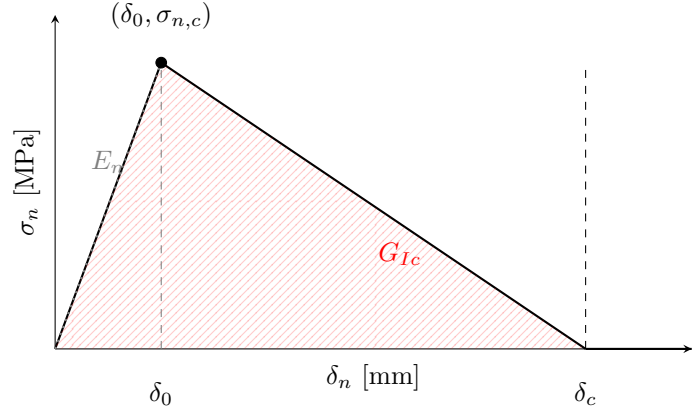

Once the initiation criterion is satisfied, damage evolves according to an energy-based softening law governed by the critical fracture energy \( G_c \). In this work, we employ a bilinear traction-separation law \cite{Ghosh2019}, as illustrated in Figure~\ref{fig:cohesive_law}. This law captures the linear elastic response up to peak strength, followed by linear softening until complete decohesion.

For the damage process to remain physically consistent, the total energy required for crack propagation (fracture toughness) must exceed the elastic strain energy stored at the moment of damage initiation. This imposes specific constraints on the cohesive parameters for standardized Mode I and Mode II fracture configurations:
\begin{align}
G_{Ic} &= \frac{1}{2} \sigma_{n,c} \delta_{n,c} > \frac{1}{2} \sigma_{n,c} \delta_{n,0} = \frac{\sigma_{n,c}^2}{2E_n}, \label{contrainte DCB} \\
G_{IIc} &= \frac{1}{2} \sigma_{t,c} \delta_{t,c} > \frac{1}{2} \sigma_{t,c} \delta_{t,0} = \frac{\sigma_{t,c}^2}{2E_t}. \label{contrainte ENF}
\end{align}
Here, $G_{Ic}$ and $G_{IIc}$ are the critical energy release rates, $\sigma_{n,c}$ and $\sigma_{t,c}$ are the peak cohesive strengths, and $E_n, E_t$ are the interface penalty stiffnesses. These inequalities ensure that a stable softening branch exists, allowing for progressive damage evolution.

Under mixed-mode loading conditions, the interaction between opening and shearing modes is captured using the Benzeggagh–Kenane (BK) criterion \cite{Benzeggagh1996}:
\begin{equation}
    G_c = G_{Ic} + (G_{IIc} - G_{Ic}) \left( \frac{G_{II}}{G_I + G_{II}} \right)^\eta,
\end{equation}
where $G_I$ and $G_{II}$ are the instantaneous energy release rates in Mode I and Mode II, respectively, and $\eta$ is a material parameter (typically $\eta \approx 2$ for carbon-epoxy composites \cite{AbdelMonsef2023}). This formulation ensures a smooth and phenomenologically accurate transition of fracture toughness across different mode mixities.

\subsection{Finite element implementation}
\begin{figure}[H]
    \centering
    \includegraphics[width=1\textwidth]{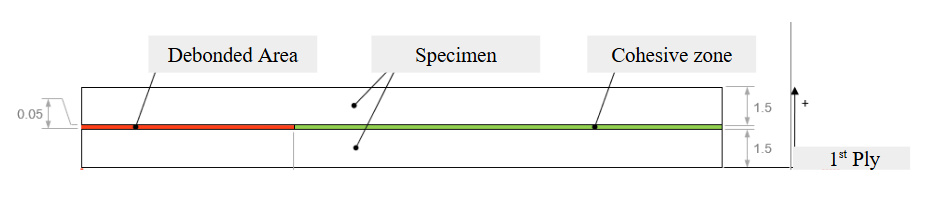}
    \caption{Cohesive zone modeling using zero-thickness interface elements (\texttt{COH3D8} in Abaqus).}
    \label{fig:spe_co}
\end{figure}

In the finite element method, cohesive elements are inserted between bulk (continuum shell or 3D) elements along potential failure interfaces to simulate interfacial debonding. These elements are typically of zero thickness and require a dedicated kinematic formulation to account for the relative displacements across the interface (see Figure~\ref{fig:spe_co}). In \textsc{Abaqus} for example, the cohesive behavior of interfaces is typically modeled using the eight-node three-dimensional cohesive element \texttt{COH3D8}, which is specifically designed to capture fracture and delamination between solid elements.

This approach enables the implementation of a traction–separation law that governs both damage initiation and damage evolution. Conventionally, numerical simulations using cohesive elements assume constant values for peak stress and fracture energy throughout the interface. However, this assumption often fails to reflect the behavior of real materials, especially composites, as it neglects spatial variability and stochastic nature of mechanical properties.

An essential parameter in cohesive‑zone modeling is the process‑zone length, \(l_{\mathrm{pz}}\). This length characterizes the region ahead of the crack tip where cohesive tractions act and where energy is dissipated through micro‑damage mechanisms. \(l_{\mathrm{pz}}\) controls the localization of damage and has a strong impact on the stability and mesh‑sensitivity of simulations that employ cohesive elements. For composite laminates, a reliable estimate of \(l_{\mathrm{pz}}\) is essential. Harper and Hallett~\cite{Harper2008} showed that neglecting this length leads to mesh-dependent predictions in delamination analyses. Their study also proposes modified expressions for the numerical cohesive‑zone length based on material and geometric parameters. For a linear cohesive law, the theoretical estimate is
\begin{equation}
l_{\mathrm{pz}} \propto \frac{E_{\mathrm{eq}}\,G_c}{\sigma_{\max}^2},
\end{equation}
where \(E_{\mathrm{eq}}\) is the equivalent elastic modulus of the composite, \(G_c\) the critical energy release rate, and \(\sigma_{\max}\) the peak cohesive strength. In our work, we conducted dedicated numerical simulations to calibrate \(l_{\mathrm{pz}}\) to considered composite materials and modeling assumptions. This calibration allowed us to choose a cohesive element length that ensures progressive and stable crack growth in our finite element models.

\section{Stochastic interface modeling using finite element method and random fields}
\label{sec:proposed_modeling}

\subsection{Random field fundamentals}

Let $D \subset \mathbb{R}^d$ be an arbitrary spatial domain (with $d=2$ for a 2D interface). A random field $H(\mathbf{x}, \omega)$ \cite{Vanmarcke1988} is defined as a collection of random variables indexed by a continuous spatial parameter $\mathbf{x} \in D$:
\begin{equation}
    H : D \times \Omega \to \mathbb{R}, \quad (\mathbf{x}, \omega) \mapsto H(\mathbf{x}, \omega),
\end{equation}
where $\Omega$ is the probability space representing the stochastic uncertainty, and $H(\mathbf{x}, \omega)$ denotes the field value at location $\mathbf{x}$ for a given realization $\omega$.

In this work, we model an interface property $\theta \in \mathbb{R}$ using an ergodic stationary Gaussian random field \cite{Wang2013}, characterized by:
\begin{itemize}
    \item \textbf{A common marginal probability distribution} $f_{\theta}(\cdot)$, which is invariant across the spatial domain $D$ due to the stationarity assumption. For each interface property, this marginal distribution is fully defined by two independent descriptors: the baseline mean $\mu$ and the standard deviation $S$ (governing the scatter magnitude), such that $H(\mathbf{x}, \omega) \sim \mathcal{N}(\mu, S^2)$ for all $\mathbf{x} \in D$.
    \item \textbf{A spatial dependence structure}, where the spatial correlation between any two distinct points in the domain is entirely governed by an isotropic correlation function $\rho(\cdot, \ell_{c})$. The independent descriptor $\ell_{c}$ represents the characteristic correlation length , which dictates the spatial decay scale of the statistical dependence.
\end{itemize}
The second-order statistics of the field are described by its covariance function $C(\mathbf{x}, \mathbf{x}')$ and the associated correlation function:
\begin{equation}
    C(\mathbf{x}, \mathbf{x}') = \text{Cov}[H(\mathbf{x}), H(\mathbf{x}')],
\end{equation}
\begin{equation}
    \rho(\|\mathbf{x} - \mathbf{x}'\|) = \frac{C(\mathbf{x}, \mathbf{x}')}{\mathrm{S}(\mathbf{x})\, \mathrm{S}(\mathbf{x}')},
    \label{eq:correlation_function}
\end{equation}
where $\mathrm{S}(\mathbf{x})$ represents the standard deviation at point $\mathbf{x}$ and $\text{Cov}$ denote the covariance operator.

The correlation function $\rho$ in Eq.~\eqref{eq:correlation_function} quantifies the statistical dependence between two points $\mathbf{x}$ and $\mathbf{x}'$. The parameter $\ell_c$, termed the \textbf{correlation length}, is critical as it controls the spatial scale of this dependence. A large $\ell_c$ implies strong correlations over long distances, resulting in a smooth field, whereas a small $\ell_c$ leads to rapid decorrelation and highly localized fluctuations.

\subsection{Proposed stochastic finite element framework}

Consider a baseline numerical model, denoted by a function $\phi$, representing a numerical model of a single test. This model maps a set of constant interfacial parameters $\theta$ (e.g., fracture energy, cohesive strength) to a global force–displacement response $Z(u) = \phi(u, \theta)$.

\begin{figure}[H]\centering\includegraphics[width=0.8\textwidth]{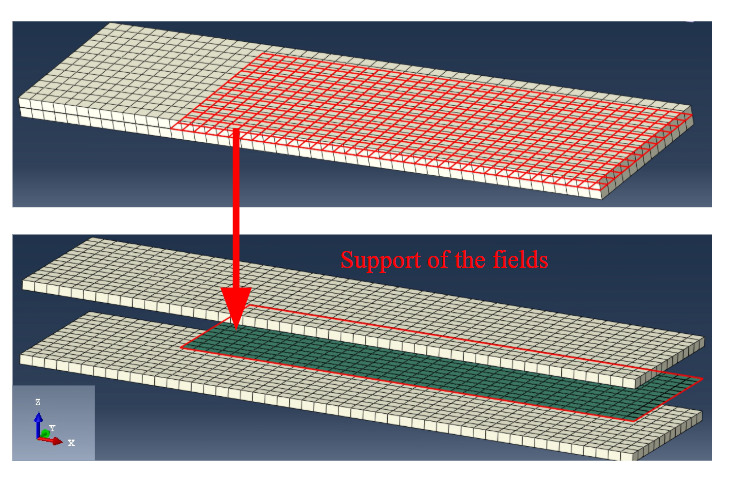}\caption{Finite element mesh serving as the spatial support $D$ for the random field.}\label{fig:support}\end{figure}

In classical finite element analysis (FEA), the interfacial parameters are assumed to be spatially constant. In our proposed stochastic framework, we replace this deterministic assumption by representing the interface properties through a spatially varying random field $\mathbf{\Theta}(\mathbf{x}, \omega)$ defined over the interface domain $D$, which coincides with the FE mesh support (Figure~\ref{fig:support}). The spatial domain is defined as $D = \{ \mathbf{x}=(x_1, x_2) \in \mathbb{R}^2 \mid 0 \leq x_1 \leq a, \, 0 \leq x_2 \leq b \}$ ($a$ and $b$ denote the physical length and width of the composite interface respectively), which is discretized using a regular mesh composed of square finite elements evaluated at $m$ discrete mesh element centers $\mathbf{x}_1, \dots, \mathbf{x}_m \in \mathbb{R}^2$. To implement this framework within a simulation-based Stochastic Finite Element Method (SFEM), each individual cohesive property $\theta_k$ undergoes spatial discretization independently. Here, $k \in \{1, 2, 3, 4\}$ denotes the four interdependent interface parameters, namely the two critical fracture energies ($G_{Ic}, G_{IIc}$) and the two peak cohesive strengths ($\sigma_{n,c}, \sigma_{t,c}$). Gathering the localized values across the mesh support for a given property yields a finite-dimensional random vector $\mathbf{\Theta}_k = [\theta_k(\mathbf{x}_1), \dots, \theta_k(\mathbf{x}_m)]^T \in \mathbb{R}^{m}$, whose multivariate normal joint probability density function $f(\mathbf{\Theta}_k)$ is expressed as:
$$f(\mathbf{\Theta}_k) = \frac{1}{(2\pi)^{m/2} \det(\mathbf{\Sigma}_k)^{1/2}} \exp\left( -\frac{1}{2} (\mathbf{\Theta}_k - \mu_{k}\mathbf{1}_m)^T \mathbf{\Sigma}_k^{-1} (\mathbf{\Theta}_k - \mu_{k}\mathbf{1}_m) \right),$$
where $\mu_{k}$ is the constant baseline mean value for the $k$-th property, $\mathbf{1}_m \in \mathbb{R}^m$ is a vector of ones, and $\mathbf{\Sigma}_k \in \mathbb{R}^{m \times m}$ is the spatial covariance matrix. The coefficients of $\mathbf{\Sigma}_k$ enforce the spatial coherence across the domain and are populated via an analytical spatial exponential covariance kernel evaluating the Euclidean distance $\|\mathbf{x}_i - \mathbf{x}_j\|$ between any two discrete mesh coordinate locations $\mathbf{x}_i$ and $\mathbf{x}_j$:
$$\Sigma_{k, ij} =S_{k}^2\rho(\|\mathbf{x}_i - \mathbf{x}_j\|, \ell_{c,k}).$$
The parameter $\ell_{c,k}$ denotes the correlation length, which governs the spatial decay scale of the statistical dependence and $S_{k}$ the constant standard deviation. In this context, isotropy means that the correlation length $\ell_{c,k}$ remains identical in all directions (horizontal or vertical) within the plane.  Although manufacturing-induced variability in composite structures can exhibit spatial anisotropy aligned with fiber orientations, an isotropic correlation length $\ell_{c,k}$ is assumed here as a baseline approximation. Consequently, the stochastic field for each individual cohesive property is fully characterized by a triad of scalar hyperparameters defined by the tuple $(\mu_{k}, S_{k}, \ell_{c,k})$, representing the mean value, the standard deviation (scatter magnitude), and the correlation length, respectively. Although a Gaussian random field can theoretically generate negative values, no such issues were encountered in this study because the dispersion around the mean remains moderate, rendering the probability of negative values occurring negligible. Furthermore, a sample filtering strategy is applied during field generation to ensure strict adherence to the physical inequalities associated with fracture mechanics. The convergence of the Finite Element model in Abaqus itself serves as an ultimate numerical safeguard, as the solver cannot execute if these criteria are violated. Nevertheless, for future developments or configurations requiring higher dispersion, the use of non-Gaussian random fields could be envisioned \cite{Grigoriu1998,Soize2006}.

To generate realizations of the discretized random vector $\mathbf{\Theta}_k$, we use a classical Monte Carlo Simulation (MCS) based on the Cholesky decomposition \cite{cressie1993statistics} of the spatial covariance matrix, defined as $\mathbf{\Sigma}_k = \mathbf{L}_k \mathbf{L}_k^T$, where $\mathbf{L}_k \in \mathbb{R}^{m \times m}$ is a lower triangular matrix. Individual sample realizations are then computed directly using the following linear transformation:$$\mathbf{\Theta}_k = \mu_{k}\mathbf{1}_m + \mathbf{L}_k \mathbf{\xi},$$where $\mathbf{\xi} \in \mathbb{R}^m$ denotes a random vector composed of independent standard normal random variables. While straightforward to implement, this direct factorization scales with an $\mathcal{O}(m^3)$ computational complexity. Consequently, when the mesh refinement parameter $m$ becomes very large, MCS encounters severe numerical bottlenecks and prohibitive memory limitations. To circumvent this curse of dimensionality and ensure numerical tractability during large-scale simulations, recourse to the Karhunen-Loève (KL) \cite{Ghanem1991} expansion is preferred. The KL framework expands the continuous random field via a truncated series of orthogonal deterministic eigenfunctions and eigenvalues, drastically compressing the stochastic space by retaining only the highest-energy modes.

This numerical stochastic framework enables  simulation of interfacial material variability, explicitly capturing how local spatial fluctuations drive the global structural response. In the specific context of composite interfaces, the correlation length represents the characteristic scale of material heterogeneity. Consequently, its accurate calibration is paramount to ensuring that the random field remains a faithful physical representation of the interface.

A key point is to differentiate the correlation length from the process‑zone length $\ell_{\mathrm{pz}}$. The former is a statistical parameter that defines the spatial extent of material fluctuations, while the latter is a deterministic physical quantity that characterizes the damage‑dissipation region inherent to the cohesive law. Keeping these two scales separate is crucial to prevent physical inconsistencies. To ensure numerical robustness, a comprehensive mesh convergence study was conducted to select an element size sufficiently fine to resolve the process zone, thereby guaranteeing stable crack propagation. Subsequently, statistical analyses were performed by varying (or holding constant) the correlation length, which remains independent of the process zone length.

\subsection{Numerical assessment of spatial variability: Random fields vs. constant property assumptions}
\label{section3.3}
The integration of spatial variability into mechanical modeling facilitates a more representative treatment of aleatory uncertainties compared to conventional deterministic methods. Rather than substituting established mechanical principles, this approach provides a statistical perspective to enhance the fidelity of numerical results by mapping manufacturing heterogeneities onto the numerical simulation. 

This assessment is done at the coupon level, using the Mode I Double Cantilever Beam (DCB) test as a representative configuration. To replicate the experimental behavior of the unidirectional laminates detailed in appendix \ref{app:experiments}, numerical models were developed in Abaqus by using three-dimensional continuum elements to represent the bulk material, while interfacial debonding was simulated by inserting zero-thickness COH3D8 cohesive elements along the failure path. These elements implement a bilinear traction-separation law that governs damage initiation and its subsequent evolution. 

To ensure a consistent comparison between modeling paradigms, both the spatially constant and spatially variable models use the same marginal probability distribution $f_{\theta}$ for the local interfacial properties. It should be noted that for this specific exploratory analysis, the marginal distribution parameters are fixed to isolate the influence of spatial correlation. The local properties are assumed to follow Gaussian distributions characterized exclusively by their respective Coefficients of Variation (CoV). Specifically, the CoVs are set to $3\%$ for $G_{I\mathrm{C}}$, $5.40\%$ for $G_{II\mathrm{C}}$, $25\%$ for $\sigma_{n,c}$, and $10\%$ for $\sigma_{t,c}$. These variation levels are directly motivated by the approximate scatter orders of magnitude observed during preliminary experimental tests and they will subsequently be calibrated using our data. Two specific experimental tests were selected at random from the five DCB samples presented in appendix \ref{app:experiments} for this comparison.

Conceptually, the constant modeling approach can be interpreted as a limiting case of the stochastic framework where the correlation length $l_{c}$ is infinite. Under this assumption of infinite persistence, parameters $\theta = (G_{Ic}, \sigma_{n,c})$ remain spatially uniform across the entire interface domain $D$, effectively neglecting localized manufacturing-induced heterogeneities. Conversely, the stochastic framework generalizes this representation by replacing the assumption of homogeneity with spatially correlated random fields $\mathbf{\Theta}(\mathbf{x}, \omega)$ governed by finite correlation lengths. For this assessment, the length $l_{c3}$ is selected, calibrated such that the correlation coefficient $\rho$ decays to $0.1$ at a distance corresponding to the third nearest neighbor ($d_{3}$) in the finite element mesh.


\begin{figure}[H]
     \centering
     \begin{subfigure}[b]{0.48\textwidth}
         \centering
         \includegraphics[width=\textwidth]{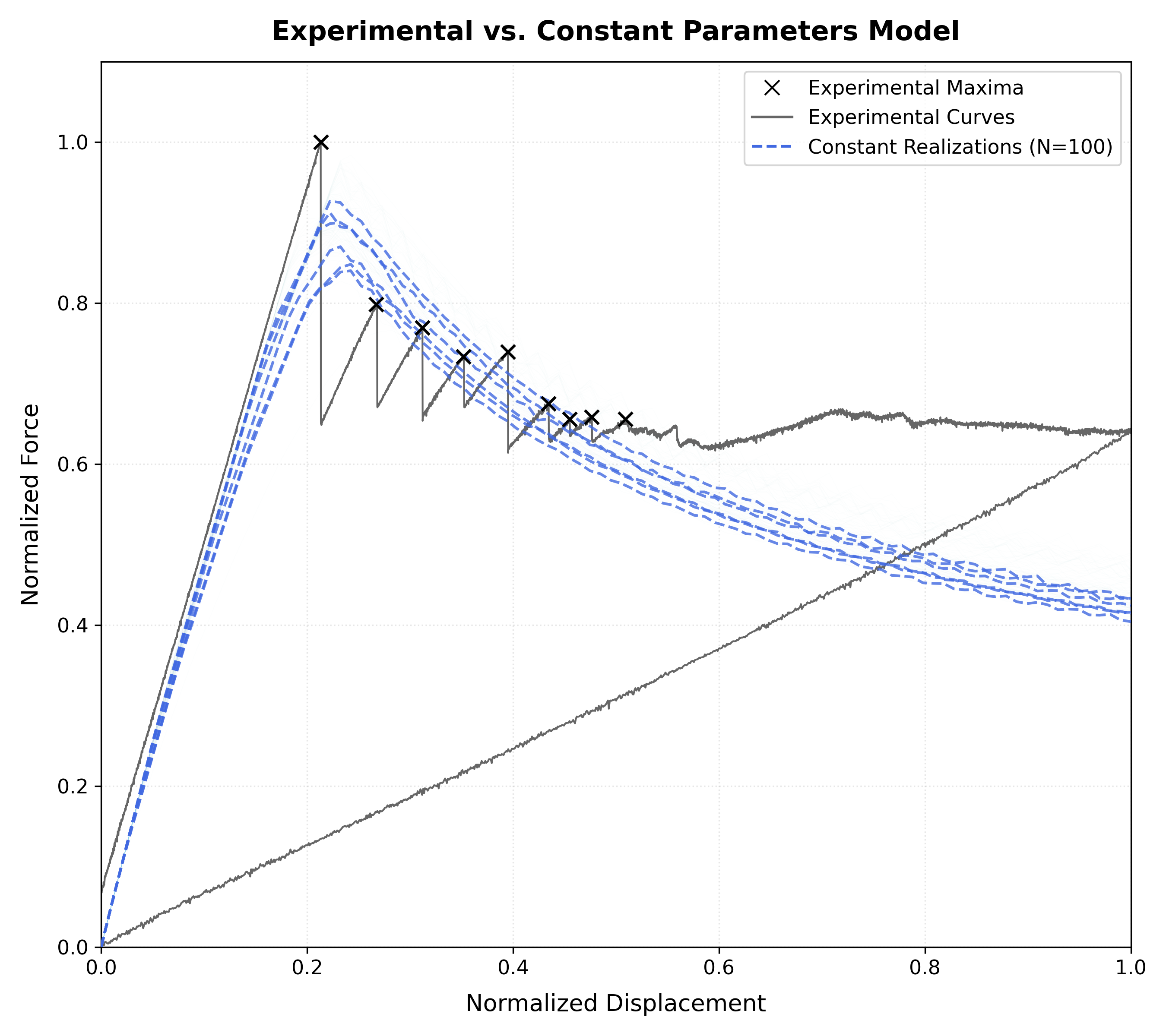}
         \caption{Constant Parameters Model}
         \label{fig:model_constant}
     \end{subfigure}
     \hfill
     \begin{subfigure}[b]{0.48\textwidth}
         \centering
         \includegraphics[width=\textwidth]{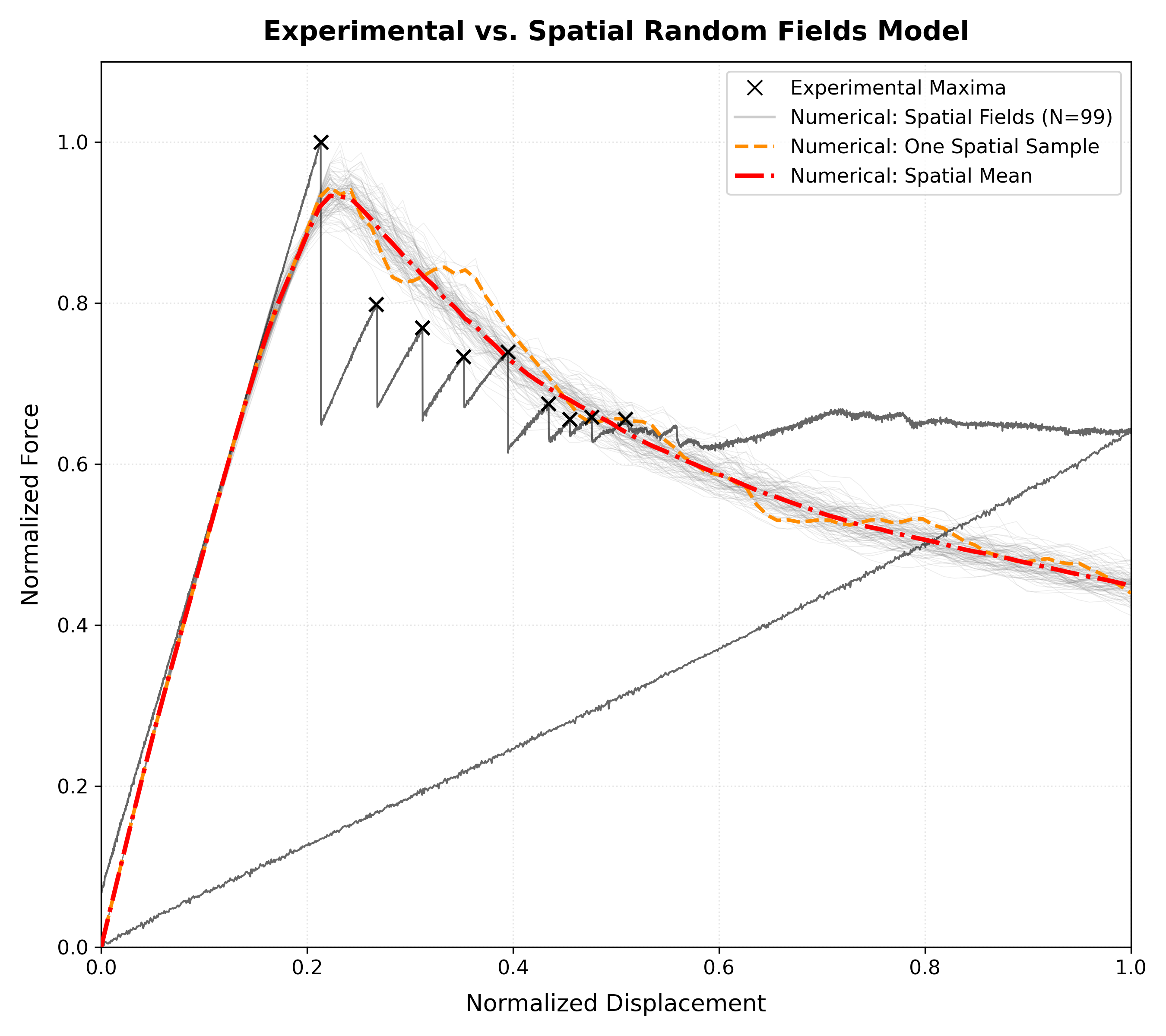}
         \caption{Spatial Random Fields Model}
         \label{fig:model_spatial}
     \end{subfigure}
     
     \caption{Comparison between the first experimental result and numerical predictions: (a) homogeneous model (with 6 numerical simulations) and (b) stochastic spatial model (with 100 numerical simulations).}
     \label{fig:comparison_final_1}
\end{figure}
\begin{figure}[H]
     \centering
     \begin{subfigure}[b]{0.48\textwidth}
         \centering
         \includegraphics[width=\textwidth]{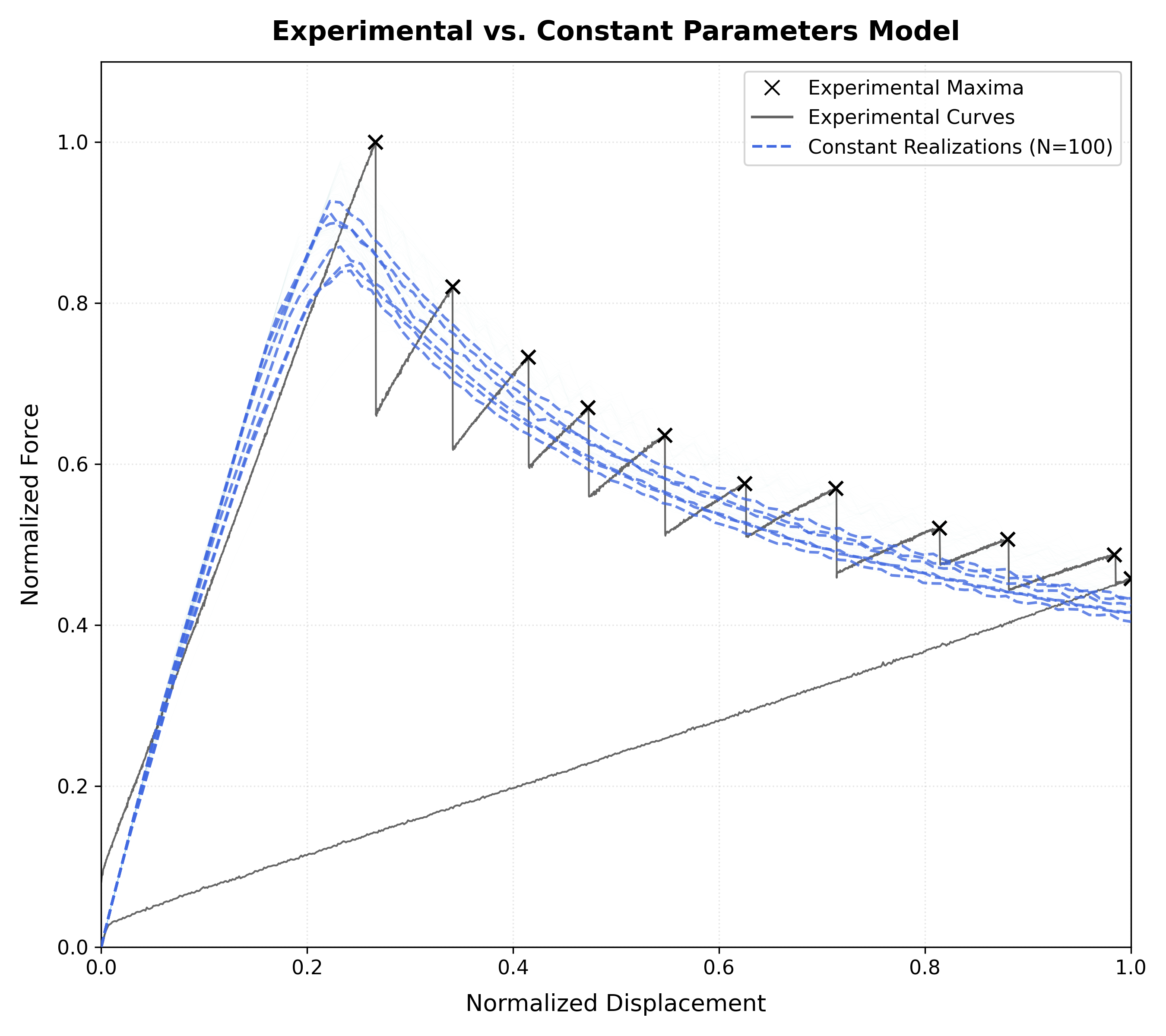}
         \caption{Constant Parameters Model}
         \label{fig:model_constant2}
     \end{subfigure}
     \hfill
     \begin{subfigure}[b]{0.48\textwidth}
         \centering
         \includegraphics[width=\textwidth]{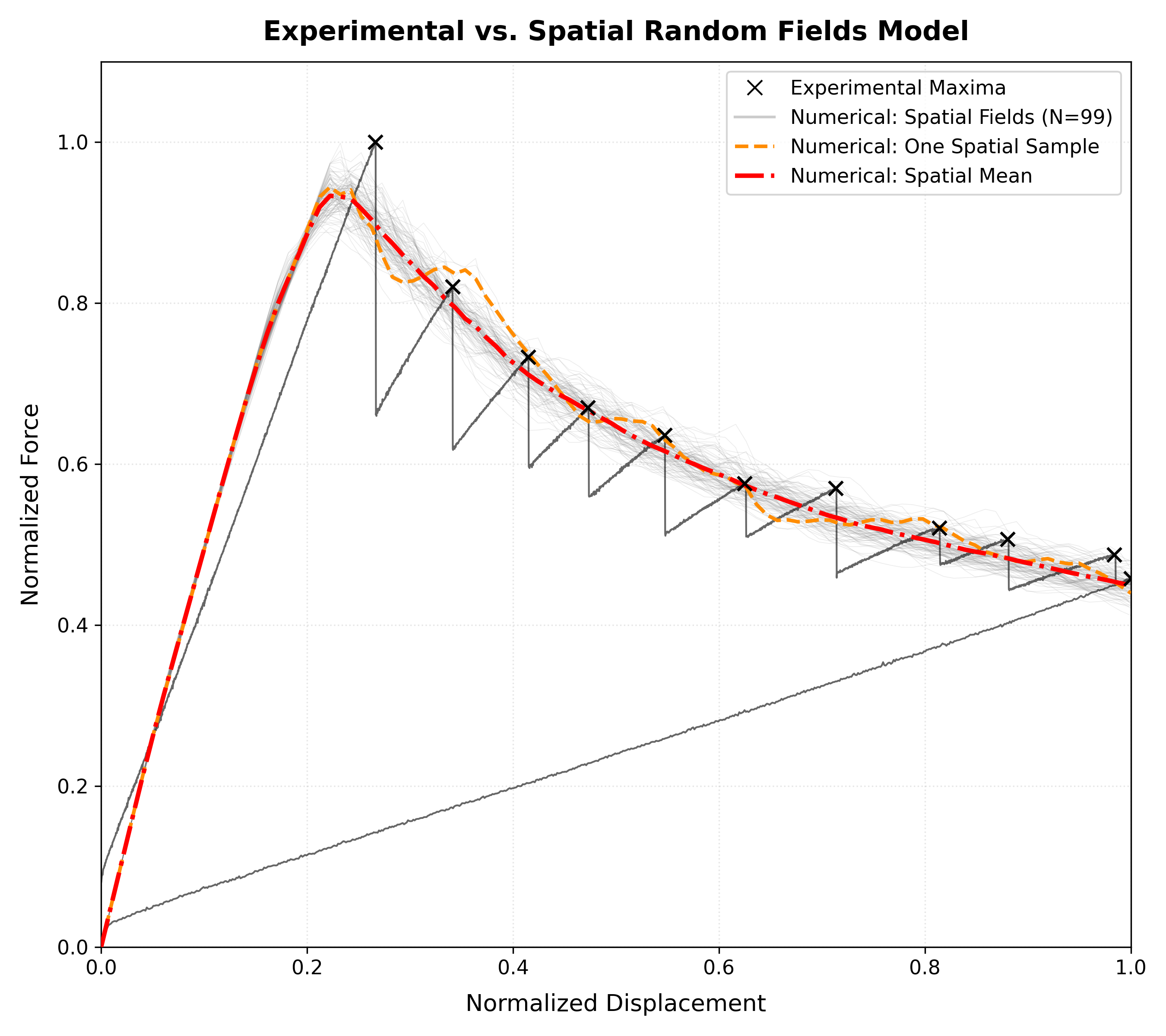}
         \caption{Spatial Random Fields Model}
         \label{fig:model_spatial2}
     \end{subfigure}
     
     \caption{Comparison between the second experimental result and numerical predictions: (a) homogeneous model (with 6 numerical simulations) and (b) stochastic spatial model (with 100 numerical simulations).}
     \label{fig:comparison_final_2}
\end{figure}
The observations derived from Figure \ref{fig:comparison_final_1} and Figure \ref{fig:comparison_final_2} underscore the qualitative differences in the resulting numerical structural responses. As illustrated by the constant realizations (blue curves in Figure \ref{fig:model_constant} and Figure \ref{fig:model_constant2}), deterministic-based models produce almost smooth force-displacement responses. While these models capture a global level of uncertainty through the variability of their values through the marginal, they are not able to reproduce the intricate R-curve behaviors and the pronounced spatial scatter observed in real composite fracture tests ~\cite{Pereira2004_ModeI,Pereira2003_ModeII,Kozik2017}. In contrast, the introduction of spatial variability through random fields yields a response that aligns more closely with experimental observations. A typical realization from this spatial model (orange curve, which is one of the curves chosen from the 100 curves shown in grey) exhibits a non-smooth softening branch with localized fluctuations (Figure \ref{fig:model_spatial} and Figure \ref{fig:model_spatial2}). These fluctuations mimic the crack propagation frequently encountered in the experimental results mentioned in appendix \ref{app:experiments}. By treating manufacturing variability as an intrinsic mechanical input via a finite correlation length $l_{c}$, the resulting spatial cloud (gray area) better reproduces the experimental dispersion than classical models, identifying statistical descriptors at the coupon scale offers a promising route for studying scale effects when transitioning from coupon  specimens to larger structural components, ensuring that local material heterogeneities are reflected in global structure.

Having established that finite spatial correlation, implemented via random fields, introduces the localized fluctuations and macroscopic dispersion inherent in composite interfaces within a numerical framework, it is now essential to systematically evaluate the influence of the underlying stochastic descriptors. From an uncertainty management perspective, this step is a prerequisite for robust modeling, as it allows us to identify the parameters to prioritize during the calibration process to ensure the accuracy of predictions. By evaluating the model's sensitivity to both the characteristic correlation length $l_{c}$ and the mathematical form of the covariance kernel, we seek to determine which parameter primarily governs the numerical macroscopic structural response. The following section presents a parametric study in which the stochastic numerical results are directly compared against the experimental datasets—specifically the Mode I Double Cantilever Beam (DCB) tests—detailed in appendix A.

\section{Sensitivity of numerical simulations to correlation length and correlation function}
\label{sec:result}
To quantify the impact of the spatial correlation structure on the macroscopic mechanical response of the numerical models, we conduct a numerical sensitivity analysis on both standardized Mode I and Mode II coupon configurations. This investigation focuses on two primary statistical descriptors: the characteristic correlation length $\ell_c$ and the regularity of the covariance kernel.

\subsection{Correlation lengths}

Three distinct correlation lengths—denoted $\ell_{c1}, \ell_{c2}, \text{ and } \ell_{c3}$—are defined based on the finite element mesh discretization to ensure a systematic evaluation. These lengths are calibrated such that the correlation coefficient decays to $0.1$ at distances corresponding to the first ($d_1$), second ($d_2$), and third ($d_3$) nearest neighbors, respectively:
\begin{align}
    \rho\left( d_1, \ell_{c1} \right) &= 0.1, \\
    \rho\left( d_2, \ell_{c2} \right) &= 0.1, \\
    \rho\left( d_3, \ell_{c3} \right) &= 0.1,
\end{align}
where $d_v = v \times e $ represents the Euclidean distance to the $v^\text{th}$ neighbor ($v \in \{1,2,3\}$), and $e$ denotes the mesh size of the numerical model. Assuming an isotropic spatial coherence on a regular mesh of square elements, the Euclidean distance between a finite element and its $v^\text{th}$ aligned neighbor is simplified to $d_v = v \times e$.

Critically, these values were selected to span the physical scale of the fracture process zone. The order of magnitude of the ratios of these correlation lengths to the process zone length are $\ell_{c1}/\ell_{\mathrm{pz}} \approx 0.4$, $\ell_{c2}/\ell_{\mathrm{pz}} \approx 0.85$, and $\ell_{c3}/\ell_{\mathrm{pz}} \approx 1.3$. Our experimental setup satisfies the hierarchy:
\[
\ell_{c1} < \ell_{c2} < \ell_{\mathrm{pz}} < \ell_{c3}.
\]
This configuration allows us to investigate two distinct regimes: one where material heterogeneity is finer than the damage dissipation zone ($\ell_c < \ell_{\mathrm{pz}}$), and one where heterogeneity spans larger distances than the fracture process ($\ell_c > \ell_{\mathrm{pz}}$). This distinction is essential for understanding the interplay between material variability and crack propagation stability.

\subsection{Covariance kernel models}
\label{sec:covariance_models}

The spatial dependency structure of the random field is governed by a correlation function, or kernel, denoted $\rho(d, \ell_c)$, where $d = \|\mathbf{x} - \mathbf{x}'\|$ is the distance between two points. Various kernel families exist in the literature \cite{Abrahamsen1997}, distinguished primarily by their behavior at the origin, which dictates the smoothness (differentiability) of the resulting field realizations.

In this study, we assess four standard isotropic kernels:

\begin{itemize}
    \item \textbf{Squared exponential:} Infinitely differentiable ($C^\infty$), producing extremely smooth field realizations.
    \[
    \rho(d,\ell_c) = \exp\left(-\frac{d^2}{2 \ell_c^2}\right).
    \]
    
    \item \textbf{Exponential:} Continuous but not differentiable at the origin ($C^0$). This kernel models processes with high local irregularity (roughness), similar to an Ornstein-Uhlenbeck process.
    \[
    \rho(d,\ell_c) = \exp\left(-\frac{|d|}{\ell_c}\right).
    \]
    
    \item \textbf{Matérn 3/2:} Once differentiable ($C^1$), offering a balance between the roughness of the exponential kernel and the extreme smoothness of the Gaussian kernel.
    \[
    \rho(d,\ell_c) = \left(1 + \frac{\sqrt{3} |d|}{\ell_c} \right) \exp\left(-\frac{\sqrt{3}|d|}{\ell_c} \right).
    \]
    
    \item \textbf{Matérn 5/2:} Twice differentiable ($C^2$), often preferred in physical modeling as it allows for smoother variations than Matérn 3/2 while avoiding the sometimes unrealistic analytic smoothness of the squared exponential.
    \[
    \rho(d,\ell_c) = \left(1 + \frac{\sqrt{5} |d|}{\ell_c} + \frac{5d^2}{3\ell_c^2} \right) \exp\left(-\frac{\sqrt{5}|d|}{\ell_c} \right).
    \]
\end{itemize}

\begin{figure}[H]
    \centering
    \includegraphics[width=1.25\textwidth]{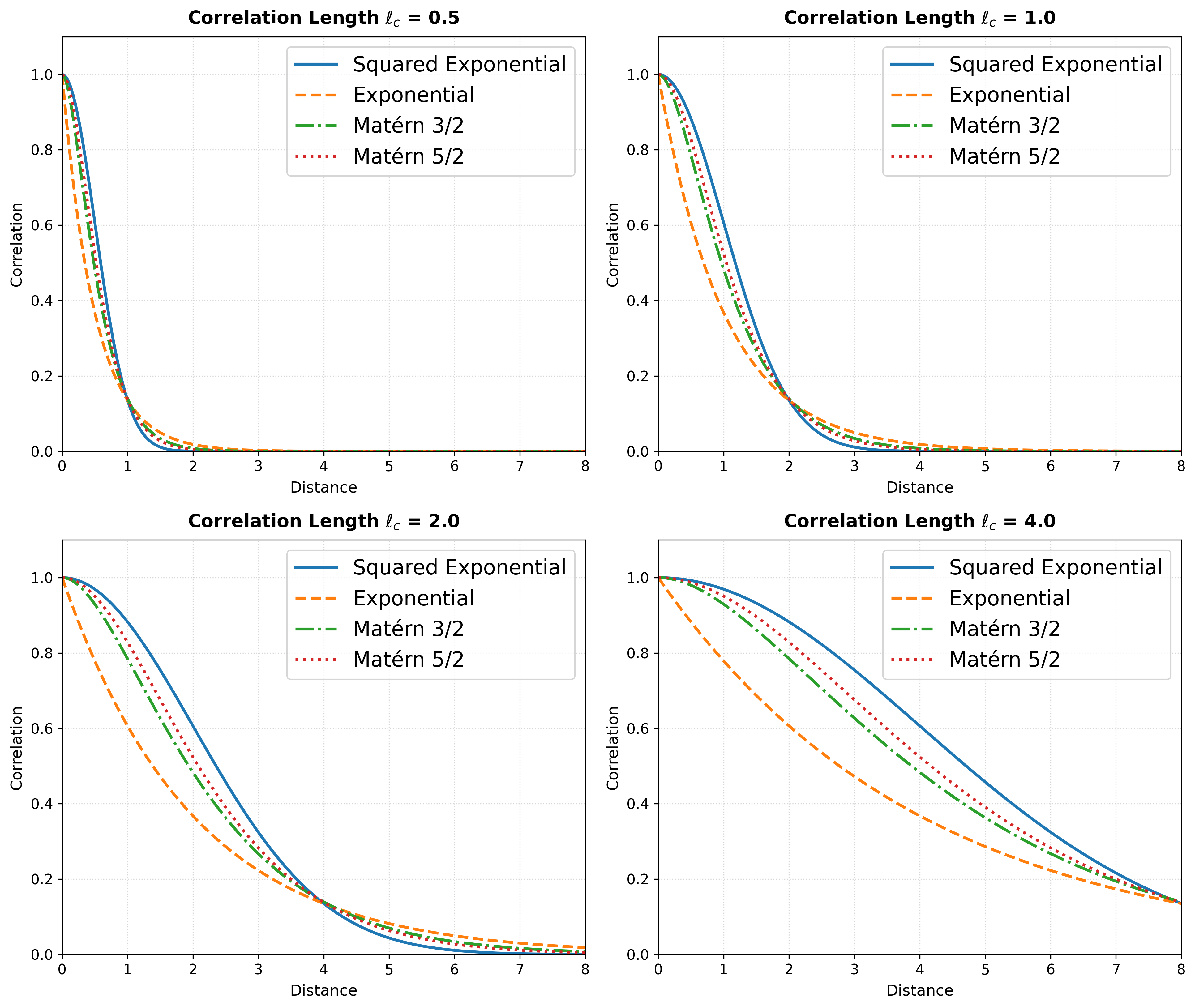}
    \caption{Influence of the correlation length $\ell_{c}$ on the decay profile of the spatial correlation function.}
    \label{fig:correlation}
\end{figure}

As illustrated in Figure~\ref{fig:correlation}, the correlation length $\ell_c$ is the parameter controlling the rate of decay of statistical dependence. Physically, a small $\ell_c$ implies rapid decorrelation, representing material properties that fluctuate over short distances (high heterogeneity). Conversely, a large $\ell_c$ indicates strong spatial persistence, corresponding to properties that vary gradually across the interface. This parameter thus serves as a bridge between the scale of microstructural heterogeneities and the macroscopic continuity of the mechanical response.

The subsequent analysis investigates the sensitivity of the numerical model to these four correlation kernels across the defined range of correlation lengths. By comparing the stochastic simulation results with experimental data, we aim to identify the kernel and length scale that most accurately capture the spatial variability inherent to the composite interface properties.

\subsection{Parametric study results}

The stochastic cohesive zone model is governed by a set of fracture energies ($G_{Ic}, G_{IIc}$) and cohesive strengths ($\sigma_{n,c}, \sigma_{t,c}$). These parameters are physically coupled through admissibility conditions derived from crack propagation criteria, as defined in Eqs.~\eqref{contrainte DCB} and \eqref{contrainte ENF}. These inequalities enforce a necessary consistency between energy-based and strength-based parameters, reflecting the underlying fracture mechanics.

From a statistical modeling standpoint, generating spatially correlated random fields for multiple interdependent variables presents significant complexity. To address this while ensuring physical relevance, we adopted a strategy of generating the stress- and energy-related fields \textbf{independently}, subsequently filtering samples to satisfy the physical inequalities in Eqs.~\eqref{contrainte DCB} and \eqref{contrainte ENF}. In practice, the variability ranges for these input parameters were defined such that \textbf{independent} sampling remains \textbf{positive} and \textbf{compatible} with the physical requirements (Eqs.~\eqref{contrainte DCB} and \eqref{contrainte ENF}). This approach provides a robust balance between computational efficiency and adherence to fracture mechanics principles. This verification step acts as a numerical safeguard: the Finite Element model is inherently unable to run or converge if these physical criteria are violated.

\begin{figure}[H]
    \centering
    \begin{subfigure}[t]{1\textwidth}
        \centering
        \includegraphics[width=1.2\linewidth]{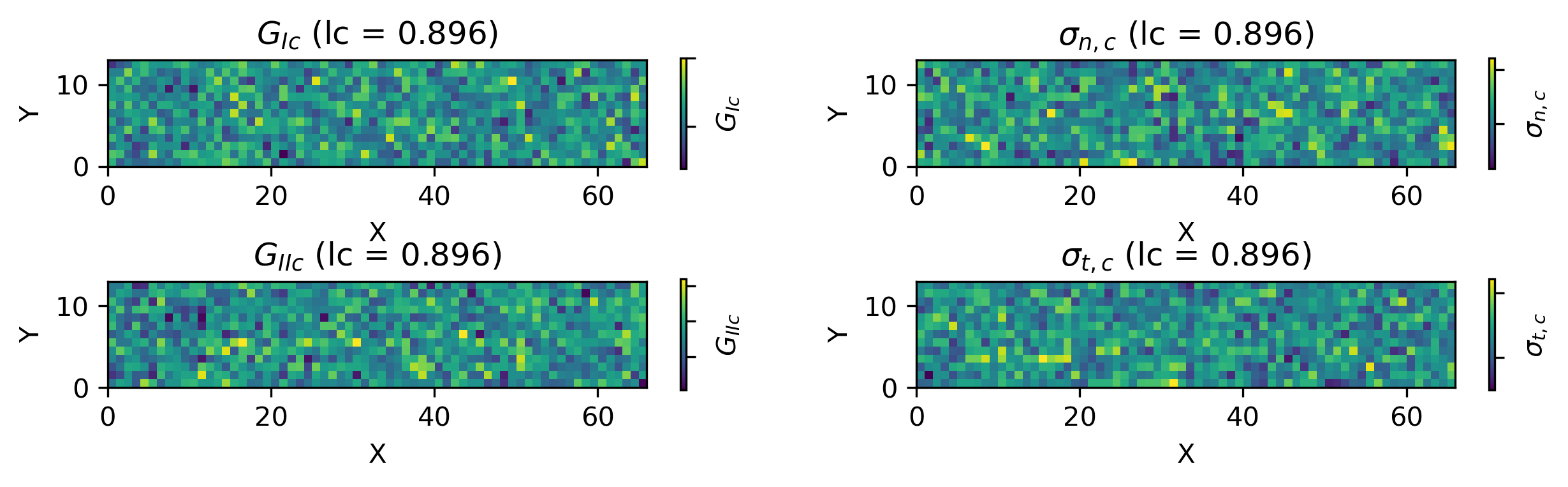}
        \caption{$\ell_{c1}$}
        \label{fig:Gaussian1}
    \end{subfigure}
    \vspace{0.5em}
    \begin{subfigure}[t]{1\textwidth}
        \centering
        \includegraphics[width=1.2\linewidth]{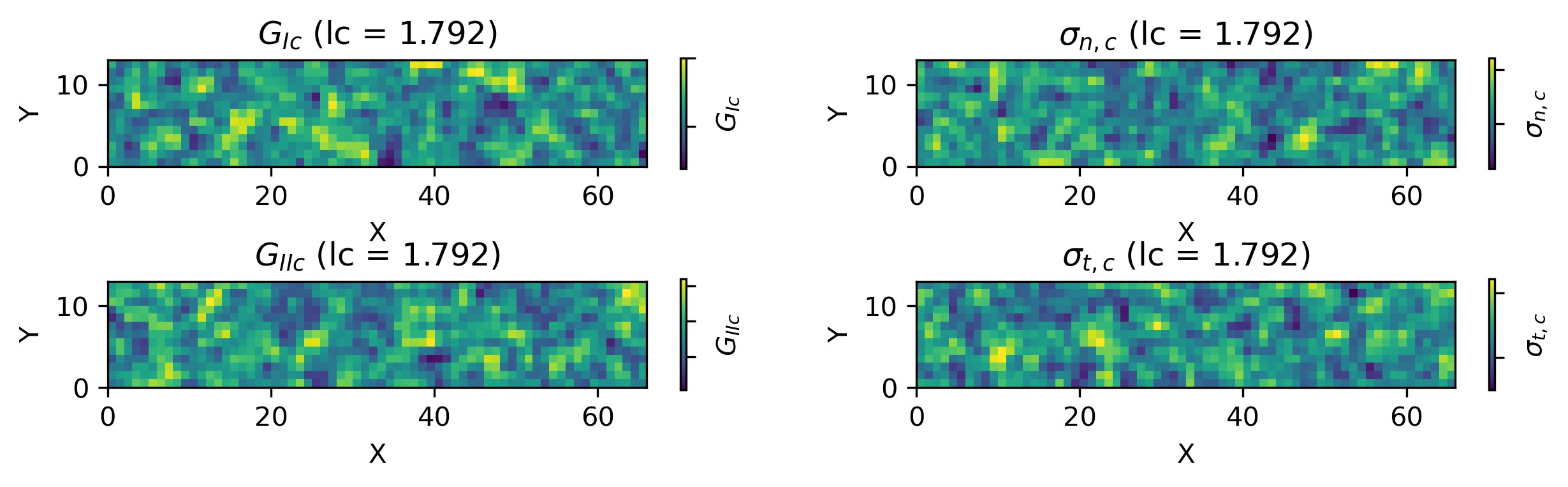}
        \caption{$\ell_{c2}$}
        \label{fig:Gaussian2}
    \end{subfigure}
    \vspace{0.5em}
    \begin{subfigure}[t]{1\textwidth}
        \centering
        \includegraphics[width=1.2\linewidth]{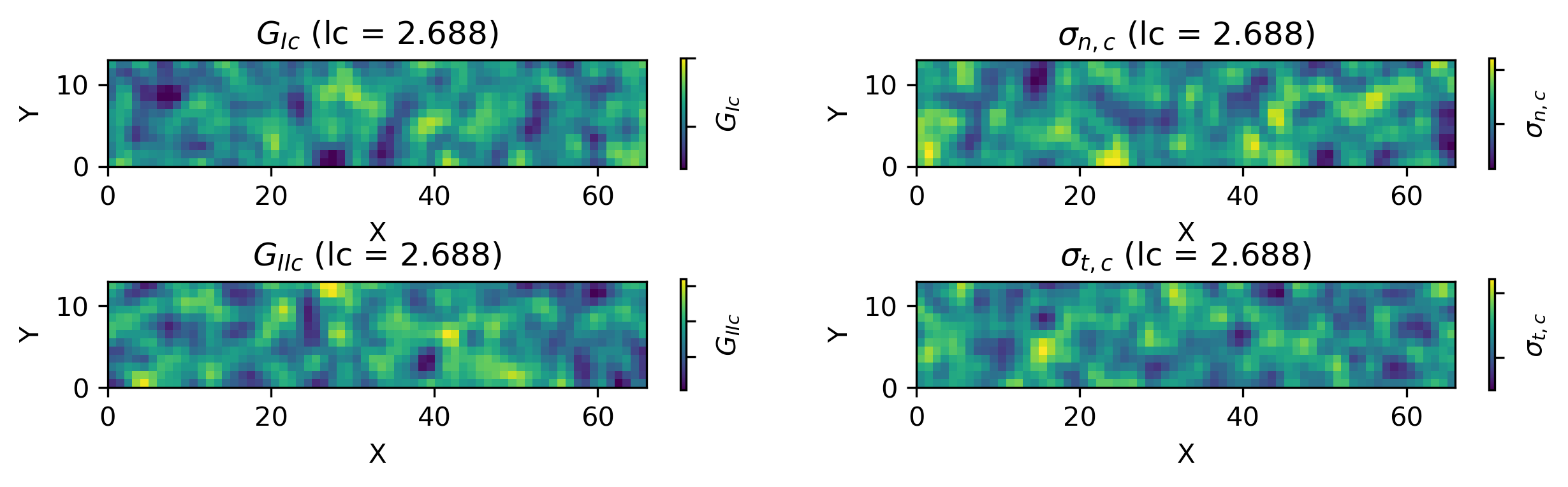}
        \caption{$\ell_{c3}$}
        \label{fig:Gaussian3}
    \end{subfigure}
    \caption{Realizations of random fields generated using the squared exponential correlation function for three distinct correlation lengths.}
    \label{fig:Gaussian_3x1}
\end{figure}

To ensure consistency across the numerical investigations, the previously defined CoV of stochastic parameters are preserved throughout this sensitivity study (from Section \ref{section3.3}). Figure~\ref{fig:Gaussian_3x1} presents typical realizations of the stochastic fields for the four cohesive parameters ($G_{Ic}, G_{IIc}, \sigma_n, \sigma_t$), generated using a squared exponential covariance kernel across three distinct correlation lengths ($ \ell_{c1}$, $\ell_{c2}$ and $\ell_{c3}$). A qualitative inspection indicates the defining role of the correlation length in shaping the spatial structure: as $\ell_c$ increases, the fields exhibit a stronger local correlation between neighboring points.

For completeness, realizations generated using alternative correlation functions—specifically the exponential and Matérn (3/2 and 5/2) kernels—are provided in appendix \ref{Weigh_IS}. While the differentiability properties of the kernels affect the local regularity (roughness) of the fields, the global trends regarding amplitude variability and field range remain consistent across kernel types. These observations underscore that, within the construction of stochastic cohesive models, the correlation length is a primary determinant of the field's topological characteristics.

The proposed stochastic modeling framework was applied at the coupon level to simulate both standardized Mode I and Mode II fracture configurations tests. For each combination of correlation length and correlation function type, a Monte Carlo simulation set of 100 field realizations was performed.

For the standardized Mode I configuration (for example, Double Cantilever Beam, DCB), the governing interface properties are \(\theta = (G_{Ic}, \sigma_{n,c})\). Figure~\ref{fig:DCB_all} illustrates the sensitivity of the macroscopic load–displacement response to the spatial correlation structure. The numerical predictions are overlaid with experimental data from five tests conducted (See appendix \ref{app:experiments}).


\begin{figure}[H]
    \centering
    
    \begin{subfigure}[b]{1\textwidth}
        \centering
        \hspace*{-2cm}
        \includegraphics[width=1.3\textwidth]{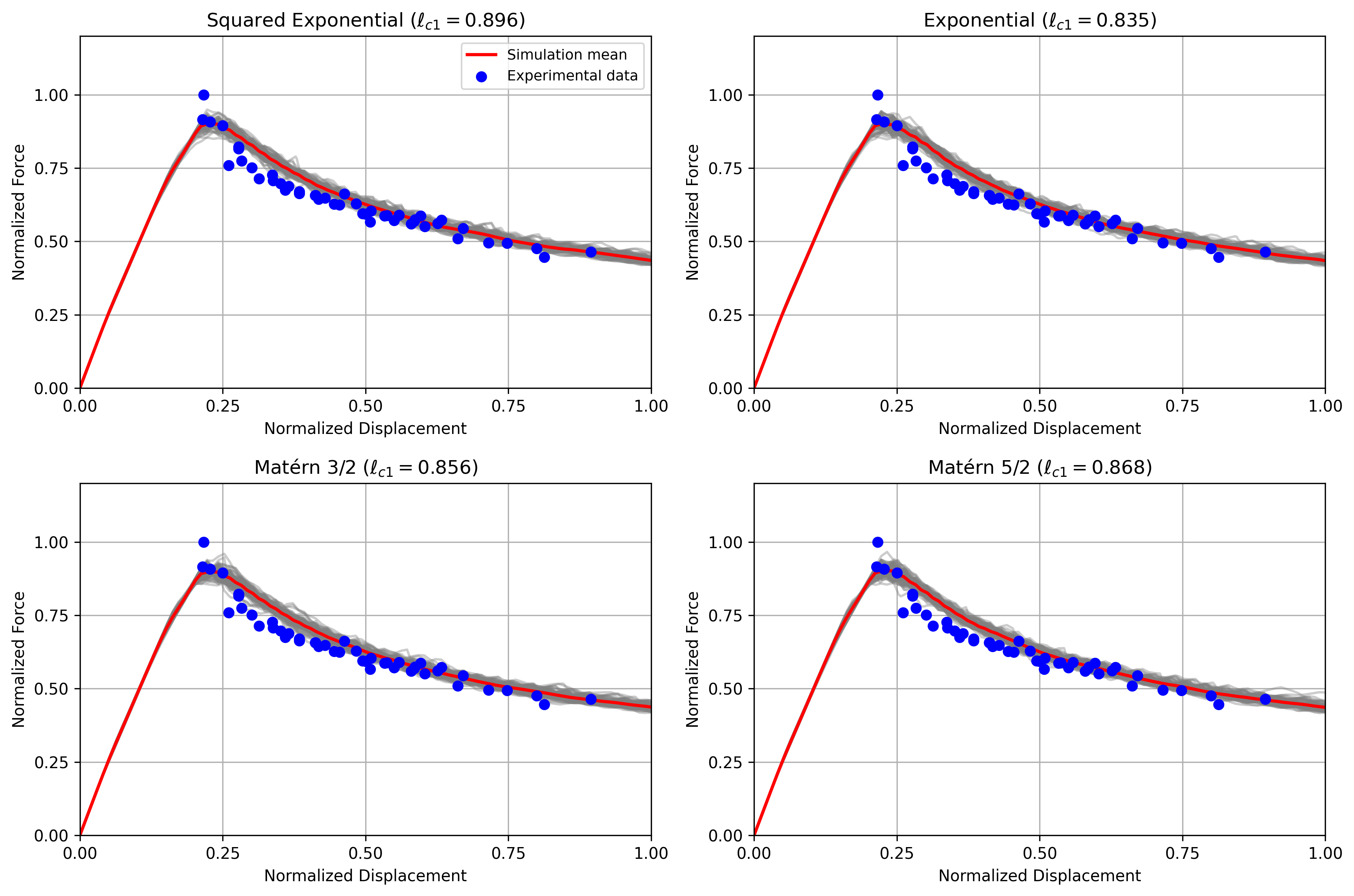}
        \caption{Correlation length $\ell_{c1}$.}
        \label{fig:DCB1}
    \end{subfigure}
\end{figure}
\begin{figure}[H]
    \ContinuedFloat
    \centering
    \begin{subfigure}[b]{1\textwidth}
        \centering
        \hspace*{-2cm}\includegraphics[width=1.3\textwidth]{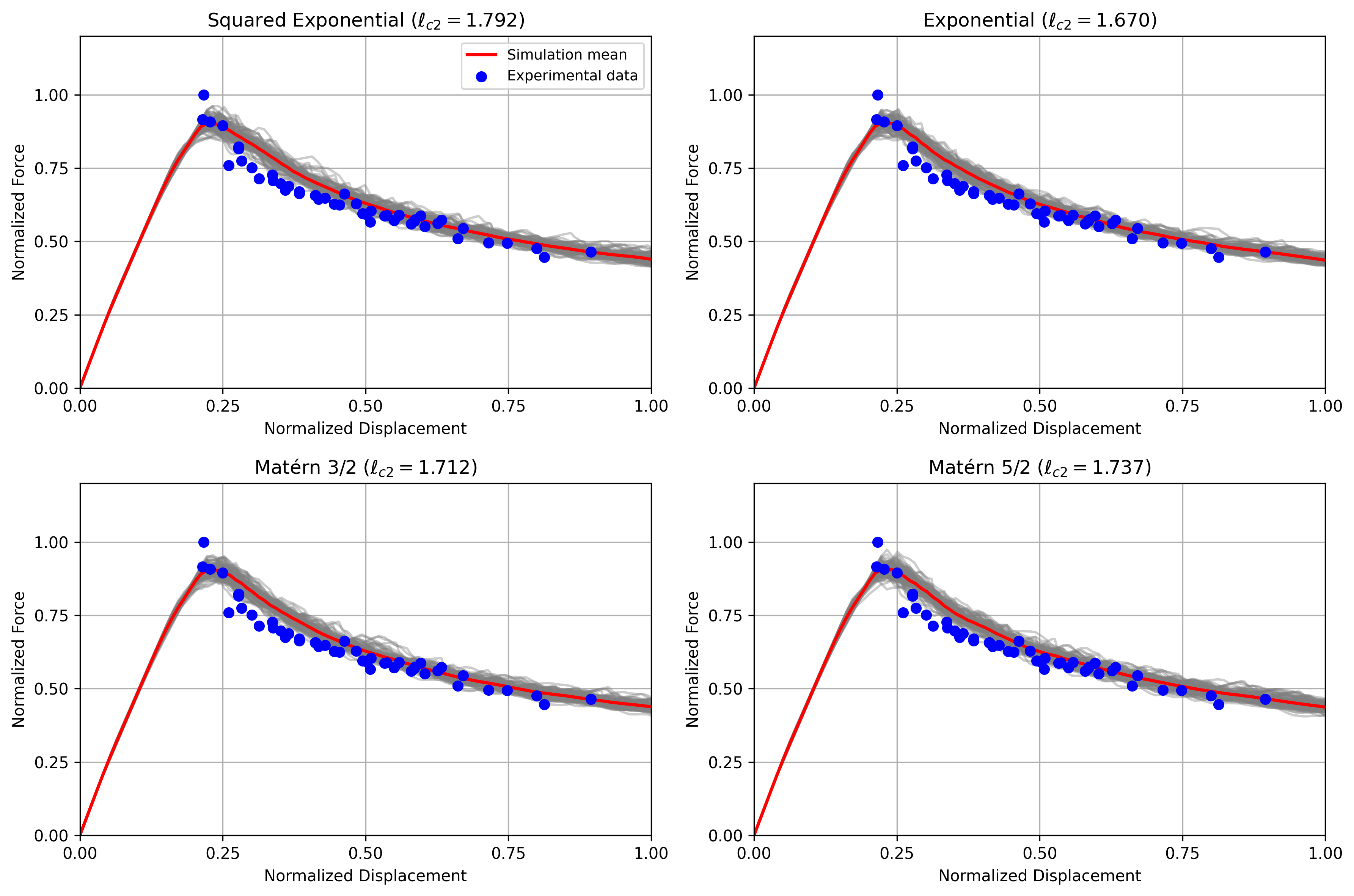}
        \caption{Correlation length $\ell_{c2}$.}
        \label{fig:DCB2}
    \end{subfigure}
\end{figure}
\begin{figure}[H]
    \ContinuedFloat
    \centering
    \begin{subfigure}[b]{1\textwidth}
        \centering
        \hspace*{-2cm}
        \includegraphics[width=1.3\textwidth]{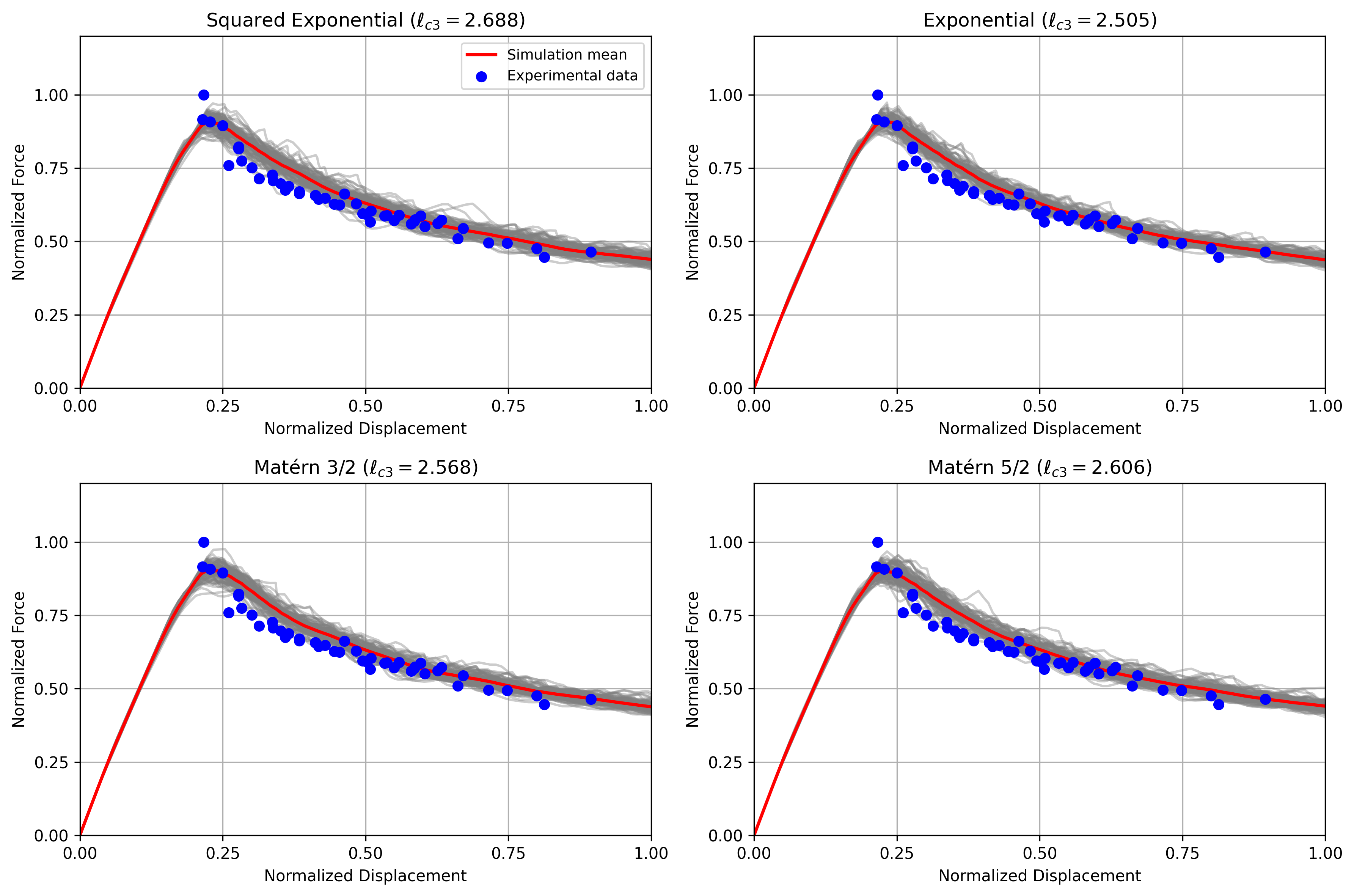}
        \caption{Correlation length $\ell_{c3}$.}
        \label{fig:DCB3}
    \end{subfigure}

    \caption{Normalized force–displacement response of DCB simulations comparing different spatial correlation functions across three correlation lengths: (a) $\ell_{c1}$, (b) $\ell_{c2}$, and (c) $\ell_{c3}$. Experimental data are shown for reference.}
    \label{fig:DCB_all}
\end{figure}

Similarly, for the standardized Mode II configuration (for example, End Notched Flexure, ENF) governed by \(\theta = (G_{IIc}, \sigma_{t,c})\), the simulation results are presented in appendix \ref{ENFTest}. In these plots, the gray curves represent the 100 stochastic realizations, the red curve denotes the ensemble mean, and the blue points correspond to the experimental data from five independent tests (See appendix \ref{app:experiments}). The results demonstrate that incorporating random fields into the cohesive zone model successfully reproduces the experimentally observed variability in a statistically consistent manner.

A comparative analysis of the results reveals two key findings:
\begin{itemize}
    \item \textbf{Influence of the kernel:} for a fixed correlation length, the four tested correlation functions produce macroscopic responses with visually similar means and dispersions. This suggests that the type of the correlation kernel has a secondary order effect on the global response of the coupon models compared to the correlation length.
    \item \textbf{Influence of the correlation length:} conversely, the correlation length significantly impacts the variance of the numerical response. This variability intensifies as the correlation length increases, a trend that remains consistent whether $\ell_c$ is smaller or larger than the characteristic length of the fracture process zone ($\ell_{pz}$).
\end{itemize}
Consequently, the correlation length $\ell_c$ emerges as the governing parameter driving the variability of the macroscopic response, whereas the specific covariance kernel plays a secondary role. This finding primarily confirms a well-known behavior extensively reported in the stochastic mechanics literature, extending its validity to the specific context of zero-thickness cohesive zone modeling. From a practical standpoint, the dominance of $\ell_c$ simplifies the inverse problem and the subsequent computational effort by restricting the calibration priority to the spatial scale of heterogeneity.

\section{Bayesian identification of random-field parameters}
\label{sec:bayesian}
Based on the sensitivity analysis of the previous section, we adopt a modeling strategy characterized by two main components: the marginal distribution $f_\theta$ and the spatial correlation length $\ell_c$. We assume that $f_\theta$ follows a Gaussian distribution $\mathcal{N}(\mu_\theta, \mathrm{S}_\theta)$ and, given the negligible influence of the correlation function, we fix the covariance kernel to the exponential form. Consequently, the identification problem reduces to estimating the hyperparameter vector $\Theta$:
\begin{equation}
    \Theta = (\mu_\theta, \mathrm{S}_\theta, \ell_c),
\end{equation}
where $\mu_\theta$ is the mean, $\mathrm{S}_\theta$ is the standard deviation, and $\ell_c$ is the correlation length.

\begin{figure}[H]
    \centering
    \begin{tikzpicture}[
        node distance=1.5cm and 1.5cm,
        block/.style={
            rectangle, draw=black, rounded corners,
            minimum height=1.4cm, minimum width=3.5cm,
            fill=blue!10, font=\small, align=center
        },
        label/.style={font=\footnotesize, align=center},
        arrow/.style={-{Latex}, thick},
    ]

    \node[label] (in1) {Input:\\ $\Theta = (\mu_{\theta}, \mathrm{S}_{\theta}, \ell_{c})$};
    \node[block, right=of in1] (model) {Numerical Model \\ $\phi(u, \Theta)$};
    \node[label, right=of model] (out1) {Macroscopic Response: \\ $Z(u)$ (Force-Displacement)};

    \draw[arrow] (in1) -- (model);
    \draw[arrow] (model) -- (out1);

    \end{tikzpicture}
    \caption{Schematic of the forward cohesive model $\phi$ mapping random-field parameters to the macroscopic force-displacement response.}
    \label{fig:forwardModel}
\end{figure}
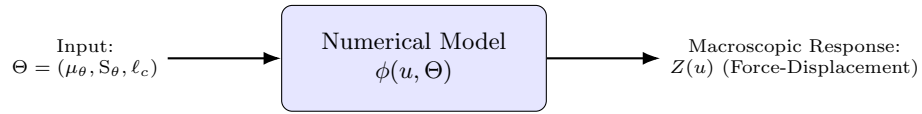

Our objective is to estimate the joint posterior distribution of $\Theta$ using a Bayesian framework. This approach allows us to reconstruct local cohesive properties from macroscopic experimental data while explicitly quantifying the uncertainties associated with the identification of $\Theta$. It serves as a mapping function generating a random field realization from $\Theta$ and propagates on numerical model $\phi$ to compute the macroscopic force-displacement response $Z(u) = \phi\!\left( u,\; \Theta \right)$.
Here, $Z(u)$ serves as the indirect output used to infer the underlying, unobservable properties of the local random field defined over the spatial domain. The overall simulation process is illustrated in Figure~\ref{fig:forwardModel}.
\subsection{Bayesian formulation of the inverse problem}

Let $Z_{\mathrm{obs}}(u)$ (with $u \in \mathbb{R}$) be the experimental observation random vector, representing the measured force-displacement response during the test. The identification problem is framed within the setting of the Bayesian approach \cite{Tarantola2005, Kaipio2005}. The objective is to update the prior knowledge about the hyperparameters $\Theta$ using the information contained in the observation. By applying Bayes' theorem, the posterior distribution is expressed as:
\begin{equation}
    \pi(\Theta \mid Z_{\mathrm{obs}}) = \frac{\pi(Z_{\mathrm{obs}} \mid \Theta)\, \pi(\Theta)}{\pi(Z_{\mathrm{obs}})} \propto \pi(Z_{\mathrm{obs}} \mid \Theta)\,\pi(\Theta),
    \label{eq:bayesRule}
\end{equation}
where $\pi(\Theta)$ is the prior distribution—constructed based on physical constraints and expert knowledge—and $\pi(Z_{\mathrm{obs}} \mid \Theta)$ is the likelihood function.
In the specific context of stochastic fracture mechanics involving non-linear finite element models \cite{Marjoram2003}, evaluating the likelihood function explicitly is computationally prohibitive. To circumvent this, we adopt a likelihood-free approach: Approximate Bayesian Computation (ABC) \cite{Sisson2018}.

It is important to note that while this study utilizes standard coupon specimens for demonstration, the proposed Bayesian identification framework is solver-agnostic and geometry-independent. It can be readily applied to any other fracture test configuration (e.g., Mixed-Mode Bending, Edge Crack Torsion) or even structural components, provided a representative numerical model exists.

\subsection{Approximate Bayesian Computation (ABC-$k$NN)}
\label{sec:abc_knn}

We use the Approximate Bayesian Computation (ABC) framework, widely recognized as the standard for parameter inference in models with intractable likelihoods \cite{Sisson2018}. Originally developed in population genetics \cite{Pritchard1999, Beaumont2002} and successfully extended to dynamical systems \cite{Toni2009} and solid mechanics \cite{Chiachio2014}, ABC substitutes the exact likelihood evaluation with a distance-based comparison between observed data and synthetic data generated by the stochastic model.

Specifically, we implement the ABC--$k$NN algorithm which combines rejection sampling with kernel density estimation to approximate the posterior distribution. It relies on a discrepancy metric to quantify the distance between a simulated response $Z_i(u) = \phi(u, \Theta_i)$ with ($i \in \mathbb{N}$) and the experimental observation $Z_{\mathrm{obs}}(u)$. We employ the $L_2$ norm over the displacement domain:
\begin{equation}
    d_i = \left\| Z_i(u) - Z_{\mathrm{obs}}(u) \right\|_{L_2} = \left( \int_{\mathbb{R}} \left| Z_i(u) - Z_{\mathrm{obs}}(u) \right|^2 \,\mathrm{d}u \right)^{1/2},\quad i \in \mathbb{N}.
\end{equation}

The posterior density is estimated using the $k$ simulations that minimize this distance (the $k$-nearest neighbors). To enhance approximation quality and mitigate the bias associated with standard rejection sampling, we apply a kernel-based weighting strategy \cite{Beaumont2002, Blum2010}. The ABC--$k$NN estimator of the posterior distribution is given by:
\begin{equation}
    \widehat{\pi}_{\text{ABC-}k\text{NN}}(\Theta \mid Z_{\mathrm{obs}}) = \sum_{i \in \mathcal{I}_k} w_i \, K_{\mathrm{H}}\!\left( \Theta - \Theta_i \right),
    \label{eq:ABCposterior}
\end{equation}
where $\mathcal{I}_k$ denotes the set of indices of the $k$ nearest neighbors, $K_{\mathrm{H}}$ is a smoothing kernel in the parameter space and the parameters $\Theta_i$ ($i \in \mathcal{I}_k$) are sampled from the prior distribution $\pi(\Theta)$. The importance weights $w_i$ are derived using a Gaussian kernel based on the discrepancy distances, assigning higher density to simulations that closely match the experiment:
\begin{equation}
    w_i = \frac{\exp\left(-\frac{d_i^2}{2\epsilon^2}\right)}{\sum_{j \in \mathcal{I}_k} \exp\left(-\frac{d_j^2}{2\epsilon^2}\right)}, \qquad \text{with} \quad \epsilon = \max_{i \in \mathcal{I}_k} d_i.
\end{equation}
This formulation ensures a robust estimation of the posterior. The complete inference procedure is summarized in Algorithm~\ref{alg:ABC}.

\begin{algorithm}[h]
\caption{ABC-$k$NN Posterior Approximation}
\label{alg:ABC}
\begin{algorithmic}[1]
    \Require  Observation $Z_{\mathrm{obs}}(u)$, Prior $\pi(\Theta)$, number of samples $N$, parameter $k$.
    \State Sample $N$ parameter sets $\Theta_i \sim \pi(\Theta)$.
    \State Simulate responses $Z_i(u) = \phi(u, \Theta_i)$ for each $i=1,\dots,N$.
    \State Compute distances $d_i = \left\| Z_i - Z_{\mathrm{obs}} \right\|_{L_2}$ for each $i=1,\dots,N$.
    \State Select the $k$ nearest neighbors:
    \[
    \mathcal{I}_k = \left\{\, i \;\middle|\; d_i \text{ is among the } k \text{ smallest} \,\right\}.
    \]
    \State Construct the posterior approximation $\widehat{\pi}_{\text{ABC-}k\text{NN}}$ using Eq.~\eqref{eq:ABCposterior}.
    \State \textbf{Output:} Approximate posterior distribution $\widehat{\pi}_{\text{ABC-}k\text{NN}}$.
\end{algorithmic}
\end{algorithm}

While the ABC-$k$NN estimator is asymptotically consistent \cite{Biau2015}, the practical selection of the algorithm's hyperparameters—specifically the number of neighbors $k$ and $\epsilon$ remain an open challenge in the literature \cite{Marin2012}. The choice of $k$ and $\epsilon$ involve a classic bias-variance trade-off. In this work, $k$ is selected based on a convergence study to ensure the stability of the identified posterior.

\subsection{Extension to multiple experiments}
In experimental mechanics, identification is rarely based on a single test. Instead, it relies on a dataset containing multiple coupons to capture variabilities. Let $\widetilde{Z}_{\mathrm{obs}} = \{ Z_{\mathrm{obs}}^{(j)}(u) \}_{j=1}^{N_{\mathrm{test}}}$ denote the set of available experimental curves. 
To incorporate information from the entire campaign, the global posterior is obtained by averaging the individual posterior approximations:
\begin{equation}
    \widehat{\pi}_{\text{ABC-}k\text{NN}}(\Theta \mid \widetilde{Z}_{\mathrm{obs}}) = \frac{1}{N_{\mathrm{test}}} \sum_{j=1}^{N_{\mathrm{test}}} \widehat{\pi}_{\text{ABC-}k\text{NN}}(\Theta \mid Z_{\mathrm{obs}}^{(j)}).
    \label{eq:multi_posterior}
\end{equation}
This aggregation procedure naturally accounts for the variability inherent in the experimental process while enforcing a consensus on the identified random-field parameters.

\subsection{Results of Bayesian identification of random-field parameters on standardized Mode I configuration (DCB as example)}

We applied the ABC-$k$NN methodology to a standard DCB configuration (using MCS), representative of current industrial certification requirements, to identify the marginal distributions of the local parameters $\theta=(G_{Ic}, \sigma_{n,c})$ and the Mode I spatial correlation length $\ell_c^I$. The experimental dataset consists of $N_{\text{test}}=5$ independent tests, where each specimen yields a macroscopic force-displacement curve denoted as $Z_{\text{obs}}(u)$ (See appendix \ref{app:experiments}). Consequently, the inverse identification problem aims to infer the vector of hyperparameters $\Theta = (\mu_{G_{Ic}}, \mathrm{S}_{G_{Ic}}, \mu_{\sigma_{n,c}}, \mathrm{S}_{\sigma_{n,c}}, \ell_c^I)$. Uniform priors were selected for all hyperparameters.

\begin{figure}[H]
    \centering
    \hspace*{-1.5cm}    \includegraphics[height=0.53\textheight]{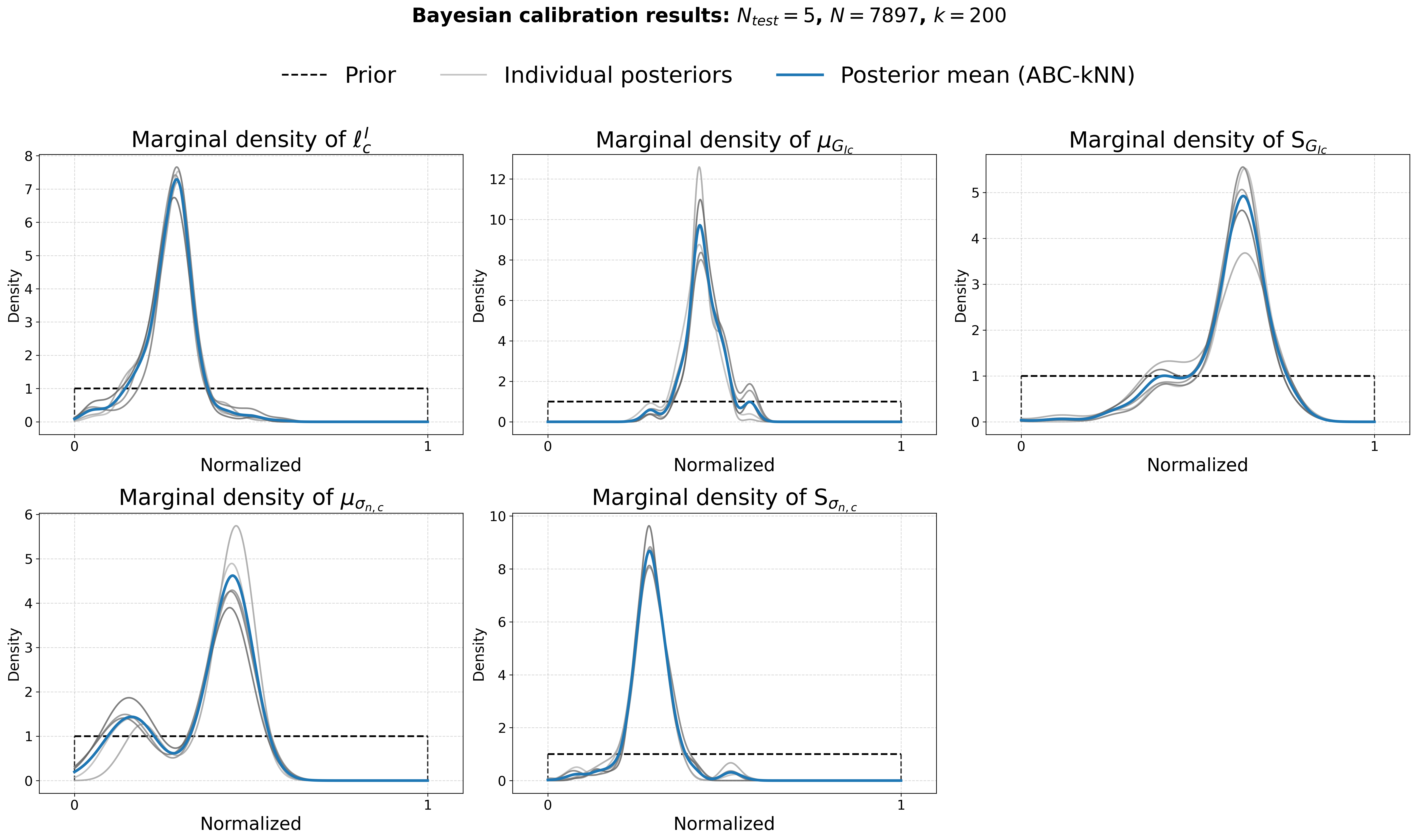}
    \caption{Posterior distributions of Mode I random field parameters calibrated from 5 DCB tests via the ABC-kNN framework.}
    \label{result:fig1}
\end{figure}

\begin{figure}[H]
    \centering
    \hspace*{-1.5cm}    \includegraphics[height=0.8\textheight]{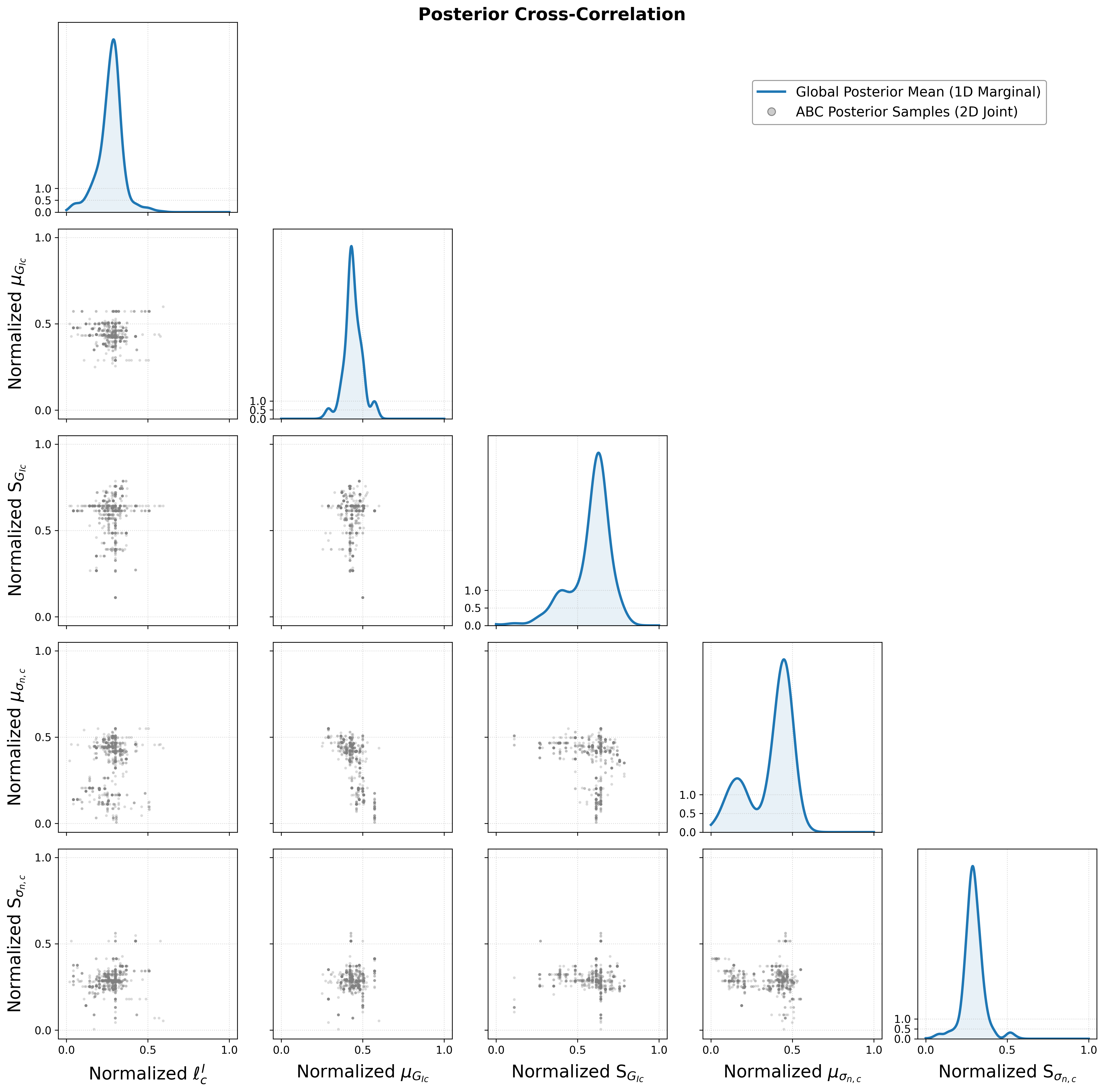}
    \caption{Posterior cross-correlation matrix for Mode I random field hyperparameters. The diagonal plots show the global 1D marginal posterior distributions, while the lower triangle scatter plots illustrate the pairwise 2D joint posterior distributions of the accepted ABC samples.}
    \label{fig:cross_correlation}
\end{figure}

The results of the Bayesian calibration are presented in Figure \ref{result:fig1}. The inference relies on a global computational budget of $N=7897$ simulations sampled from the prior. To estimate the posterior density, we utilized the $k=200$ nearest neighbors. The marginal posterior distributions (solid blue lines) exhibit a significant reduction in uncertainty compared to the uniform priors (dashed lines), indicating that the experimental data carries high informational content regarding the random field parameters. To analyze parameter dependencies, the posterior cross-correlations are examined in Figure~\ref{fig:cross_correlation}. The pairwise scatter plots generally reveal weak correlations among most hyperparameters. However, a noticeable correlation trend is observed between the mean fracture energy ($\mu_{G_{Ic}}$) and the mean cohesive strength ($\mu_{\sigma_{n,c}}$). This coupling is physically expected and completely normal, as these two parameters are intrinsically linked through the constitutive traction-separation law governing the cohesive zone interface.


\begin{figure}[H]
    \centering
    \includegraphics[height=0.5\textheight]{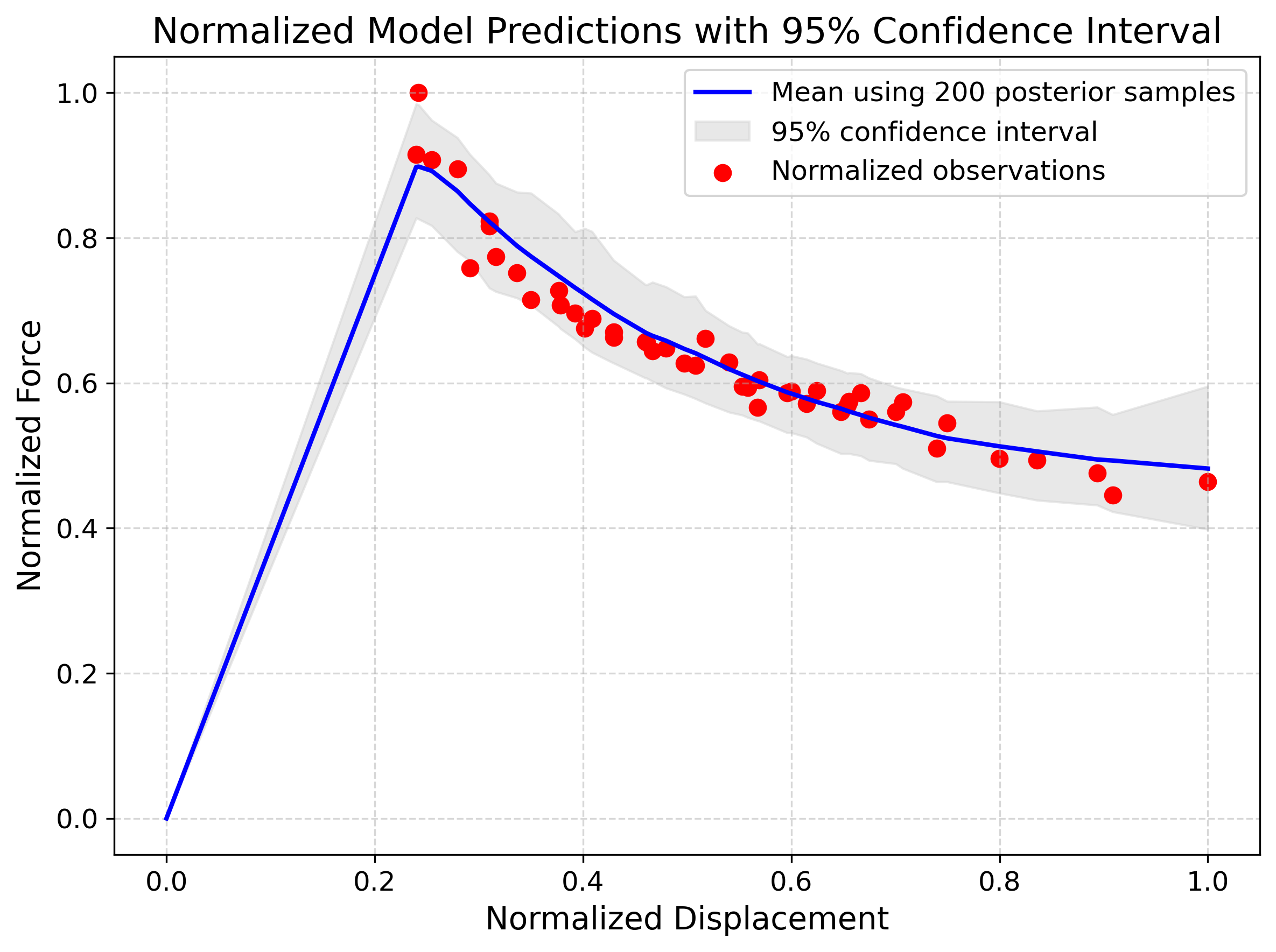}
    \caption{Comparison of the DCB response curve calibrated  with test data}
    \label{result:fig2}
\end{figure}

To validate the calibrated stochastic model, we performed a Posterior Predictive Check. Figure \ref{result:fig2} compares the normalized experimental observations (red dots) against the numerical predictions generated using samples from the identified posterior distribution. 

The blue curve represents the mean model response, while the gray shaded area corresponds to the $95\%$ confidence interval. The results demonstrate that the calibrated stochastic model effectively encompasses the experimental scatter. The $95\%$ confidence interval successfully covers the vast majority of the experimental data points, including the variability observed in the softening branch of the curve. This confirms that the identified random field parameters $(\Theta)$ and the associated correlation length $(\ell_c^I)$ accurately represent the physical variability of the interface, validating the proposed framework for virtual testing applications.

\begin{figure}[H]
    \centering
    \hspace*{-1.5cm}
    \includegraphics[height=0.5\textheight]{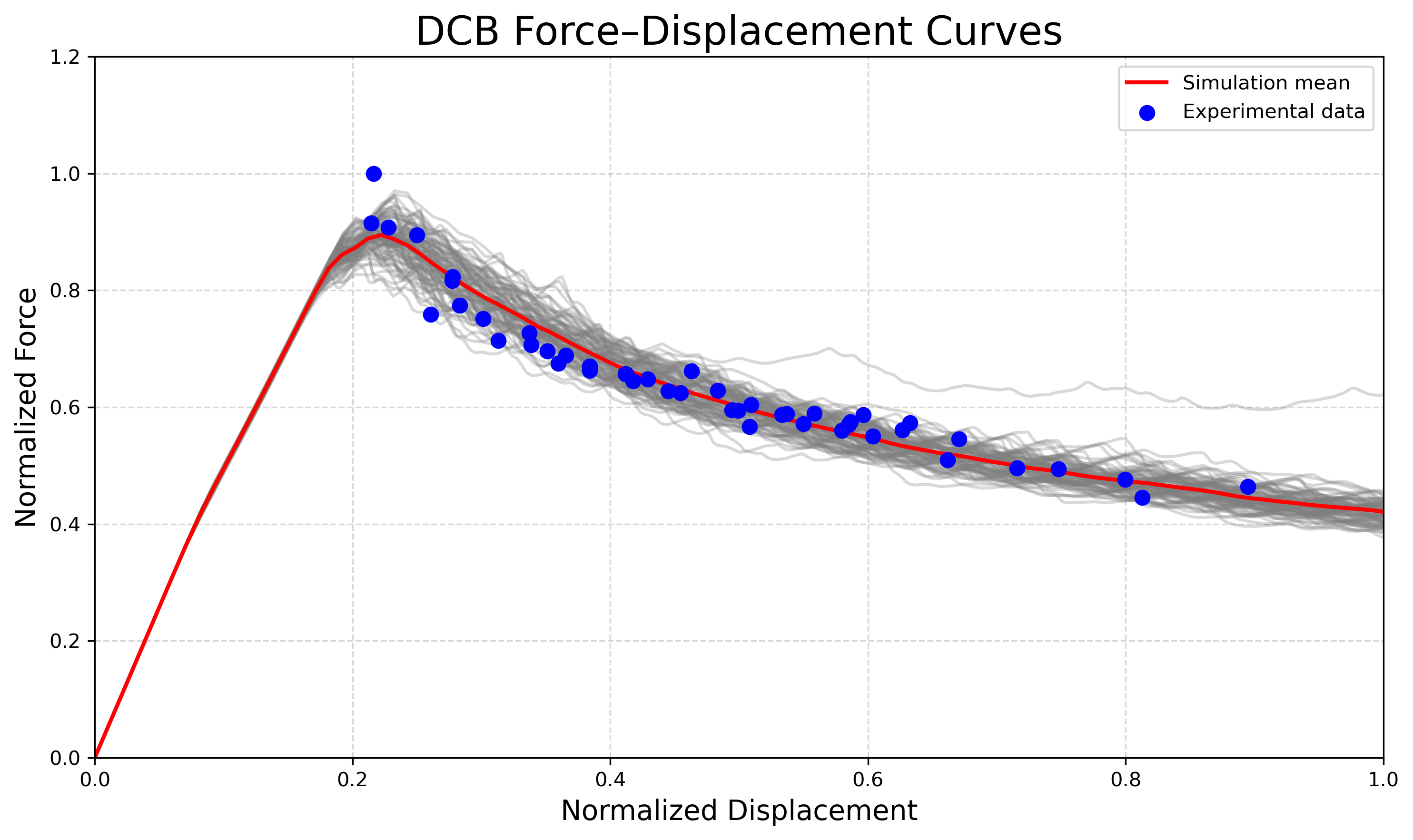}
    \caption{Forward propagation of intrinsic variability: 100 numerical realizations (grey lines) generated using the mean of the posterior distribution of $\Theta$, compared against experimental data (blue dots).}
    \label{result:fig3}
\end{figure}

To further verify the physical representativeness of the proposed framework, Figure~\ref{result:fig3} isolates the impact of spatial heterogeneity on the structural response. Unlike Figure~\ref{result:fig2}, which presented the global probabilistic confidence intervals, Figure~\ref{result:fig3} was generated using a fixed set of hyperparameters corresponding to the mean values of the identified posterior distribution ($\mu_\Theta = \mathbb{E}_{\Theta \sim \widehat{\pi}_{\text{ABC-}k\text{NN}}(\cdot \mid \widetilde{Z}_{\mathrm{obs}})}[\Theta]$ with $\mathbb{E}$ the mathematical expectation).

Using this single set of statistical descriptors, 100 independent realizations of the random field were generated and propagated through the finite element model. The resulting family of curves (shown in grey) illustrates the structural variability arising solely from the local fluctuations of the interface properties, effectively decoupling intrinsic material scatter from the epistemic uncertainty of the identification process.

It is noteworthy that with fixed hyperparameters, the model with stochastic inputs reproduces the characteristic dispersion and the non-smooth softening behavior observed in the experimental dataset (blue dots). Quantitatively, the numerical envelope encompasses 46 out of the 47 filtered experimental maxima, representing a coverage rate of approximately 97.8\%. This confirms that the identified correlation length $\ell_c$ successfully maps the specific topology of the manufacturing defects onto the macroscopic mechanical response of numerical model at coupon level.

\section{Conclusion}
\label{sec:conclusion}
This work presents a stochastic numerical framework that enables predictive treatment of uncertainties in aerospace composite structure interfaces. By replacing the assumption of a homogeneous interface with spatially correlated random fields within a finite‑element context, the approach offers a statistical methodology to translate manufacturing variability into the structural performance of bonded assemblies such as skin‑stringer joints. The comparative assessment at the coupon level demonstrates that while deterministic models with constant properties produce unphysically smooth responses, the introduction of spatial variability reproduces the non-smooth softening propagation observed in real fracture tests. A parametric study at the coupon level reveals a hierarchy of the statistical parameters: the correlation length ${\ell}_c$ emerges as the key stochastic variable governing the dispersion observed in the macroscopic response of the models, whereas the specific type of the covariance kernel has a negligible effect, greatly simplifying the stochastic modeling effort.

A major advantage of the framework is its seamless integration into existing industrial workflows. We demonstrate that ${\ell}_c$ can be identified through standard fracture tests, illustrated by the widely used Mode I (DCB) configuration in aerospace. Applying the Approximate Bayesian Computation (ABC‑$k$NN) method to this representative industrial dataset shows that uncertainty quantification can be performed at the coupon level.

Because the statistical identification relies on a limited experimental set ($N_{\text{test}}=5$), the propagation of uncertainties to larger scales must explicitly account for the epistemic uncertainty arising from this data scarcity \cite{DonfackSiewe2026_1} . Consequently, reliability assessments must incorporate both the intrinsic physical variability of the interface and the confidence bounds associated with the small sample size.

A primary perspective of this work is to leverage this stochastic framework to investigate scale effects through numerical simulation. By transitioning from the coupon scale to larger structural components, such as stiffened panels, this model will determine whether the variability observed at the coupon scale is amplified or attenuated at the structural level. 

Future work could also enhance the physical representation of the interface by introducing anisotropic correlation lengths $(l_{c}=(l_{x},l_{y}))$ aligned with fiber orientations, leveraging advanced stochastic representations of material anisotropy and structural symmetries formalized by Guilleminot and Soize \cite{guilleminot2012stochastic} in multiscale material modelingand by investigating the full propagation of combined aleatory and epistemic uncertainties from coupon tests to full‑scale structural components.

\backmatter

\section*{Declaration of competing interest}
The authors declare that they have no known competing financial interests or personal relationships that could have appeared to influence the work reported in this paper.








\begin{appendices}

\section{Experimental dataset and test results}
\label{app:experiments}

The validation of the proposed stochastic framework relies on an experimental dataset provided by Airbus, comprising standardized Mode I (DCB) and Mode II (ENF) fracture tests. The campaign involves five independent specimens for each configuration ($N_{\text{test}}=5$).

The specimens are manufactured from carbon-epoxy unidirectional (UD) laminates with a stacking sequence of $[0^\circ]_8 // [0^\circ]_8$. Each arm consists of 8 plies with a cured ply thickness (CPT) of 0.184 mm, resulting in a total specimen thickness of approximately 3 mm (see Table \ref{tab:layup}).

\begin{figure}[H]
    \centering
    \includegraphics[width=0.7\textwidth]{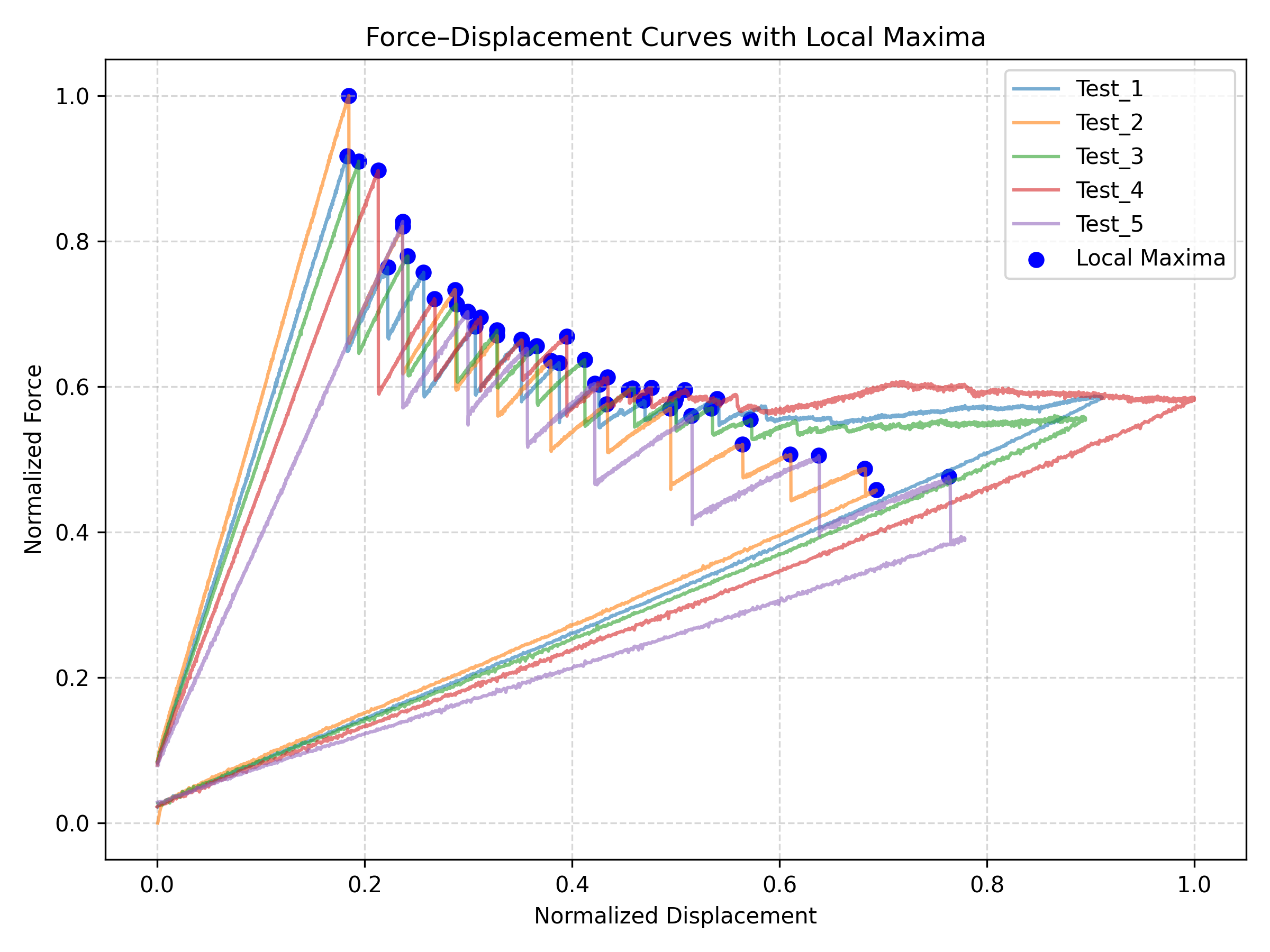}
    \caption{Normalized experimental response of the five DCB specimens (Mode I). Blue dots indicate local maxima associated with stick-slip crack propagation.}
    \label{fig:dcb_exp}
\end{figure}

\begin{figure}[H]
    \centering
    \includegraphics[width=0.7\textwidth]{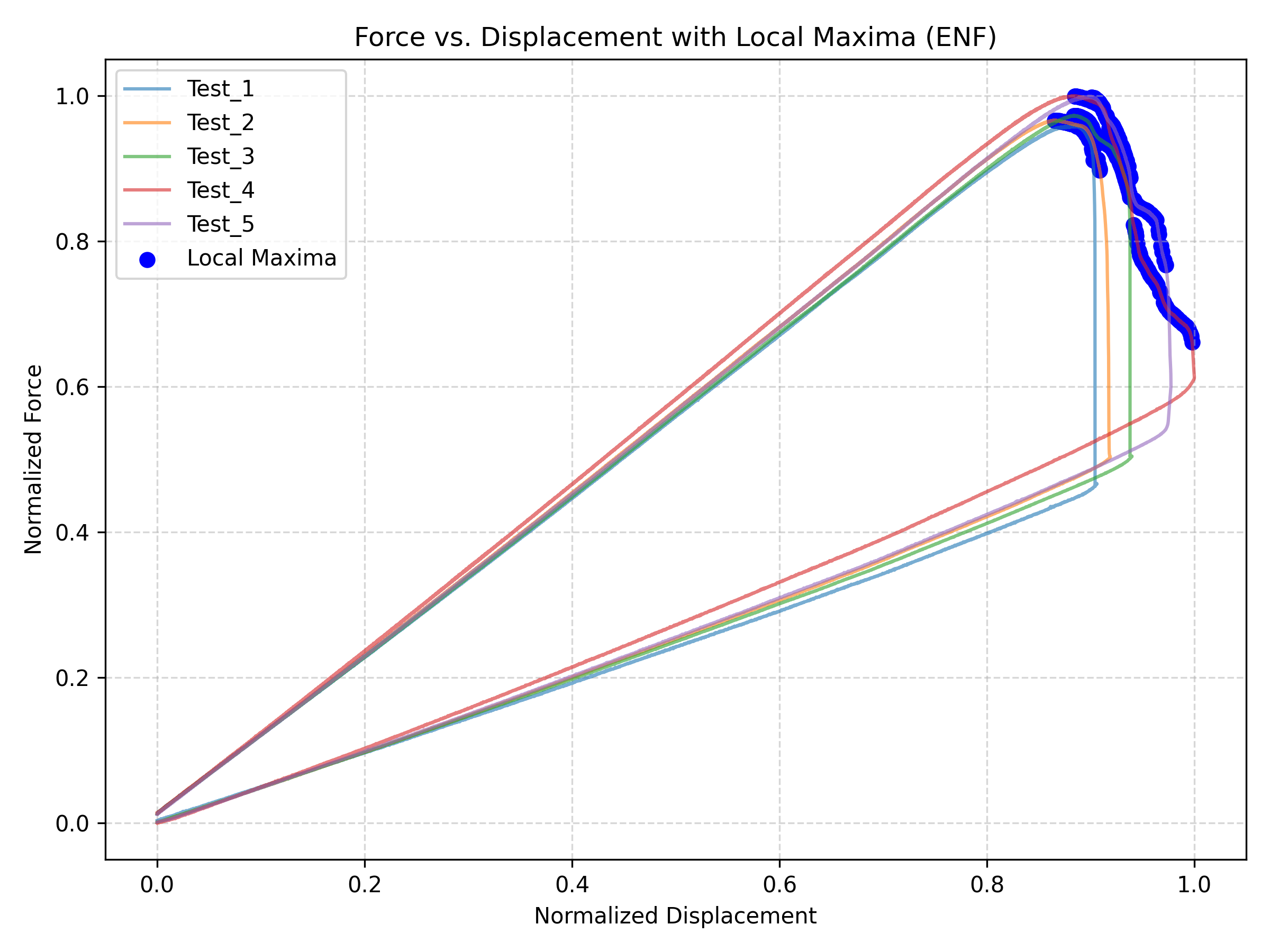}
    \caption{Normalized experimental response of the five ENF specimens (Mode II). The dispersion highlights the influence of interface heterogeneity on the macroscopic response.}
    \label{fig:enf_exp}
\end{figure}

Figures \ref{fig:dcb_exp} and \ref{fig:enf_exp} present the normalized experimental responses. 
\begin{table}[h]
    \centering
    \begin{tabular}{|c|c|c|c|}
        \hline
        \textbf{Stacking sequence} & \textbf{CPT [mm]} & \textbf{Number of plies} & \textbf{Total thickness [mm]} \\
        \hline
        $[0^\circ]_8 ~\|~ [0^\circ]_8$ & 0.184 & 16 & 2.994 \\
        \hline
    \end{tabular}
    \caption{Laminate configuration used for all DCB and ENF specimens.}
    \label{tab:layup}
\end{table}

\section{Illustration of the effect the regularity of correlation function and the effect of correlation length on random fields}
\label{Weigh_IS}

\begin{figure}[H]
    \centering
    \begin{subfigure}[t]{0.9\textwidth}
        \centering
        \includegraphics[width=\linewidth]{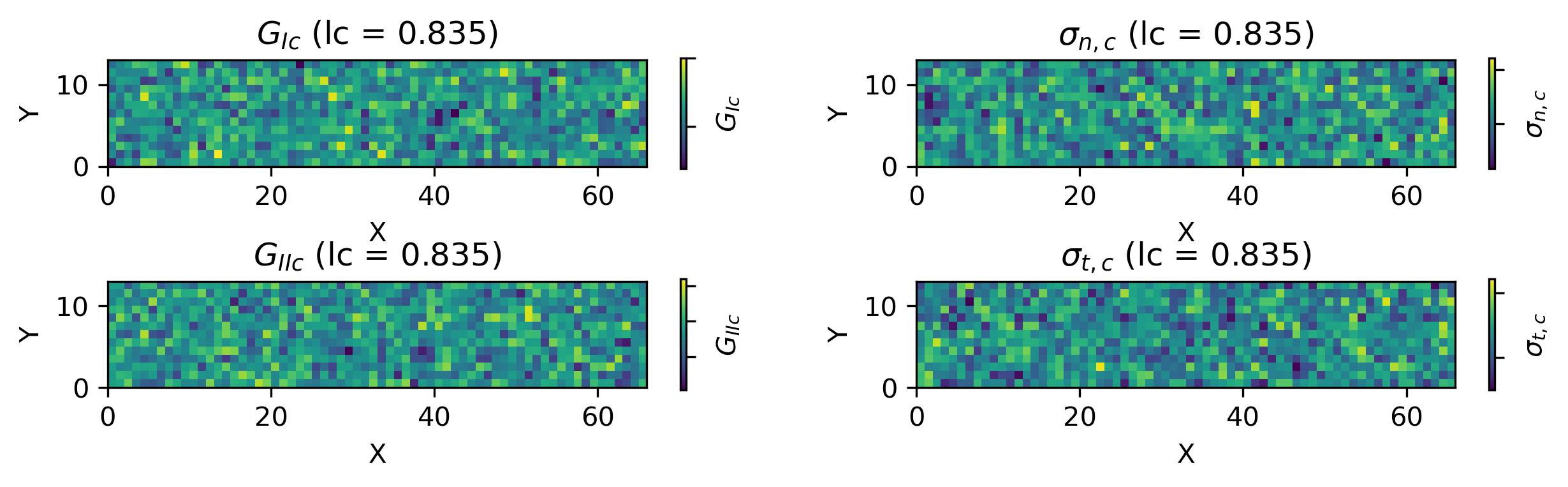}
        \caption{$\ell_{c1}$}
        \label{fig:EXP1}
    \end{subfigure}
    \vspace{0.1em}
    \begin{subfigure}[t]{0.9\textwidth}
        \centering
        \includegraphics[width=\linewidth]{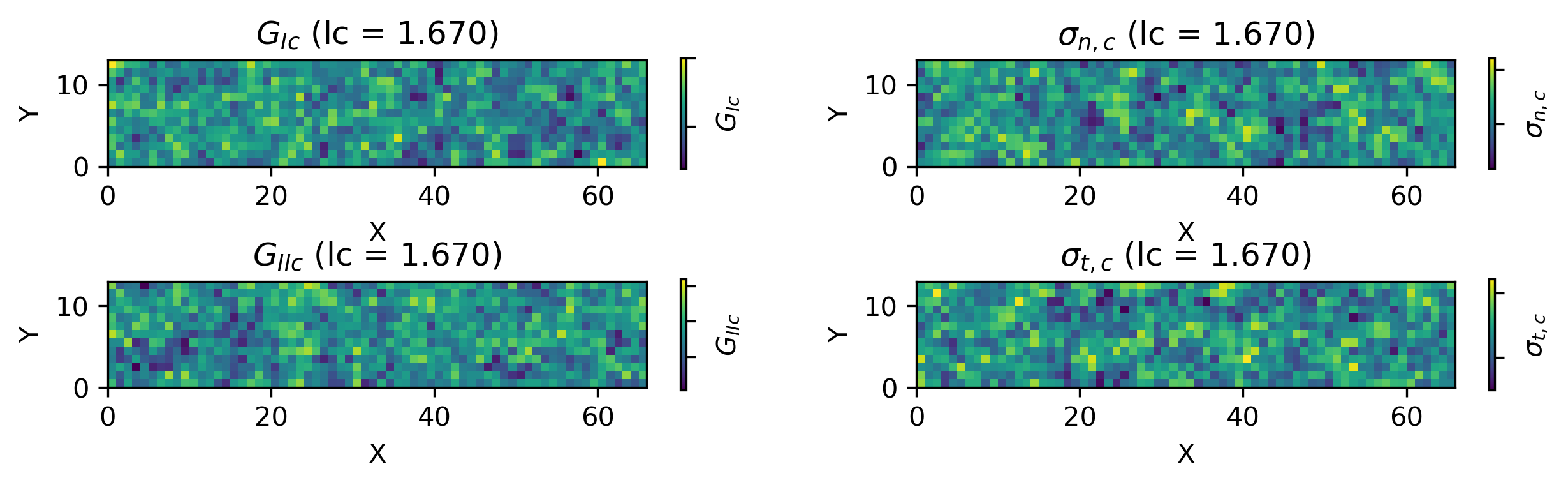}
        \caption{$\ell_{c2}$}
        \label{fig:EXP2}
    \end{subfigure}
    \vspace{0.1em}
    \begin{subfigure}[t]{0.9\textwidth}
        \centering
        \includegraphics[width=\linewidth]{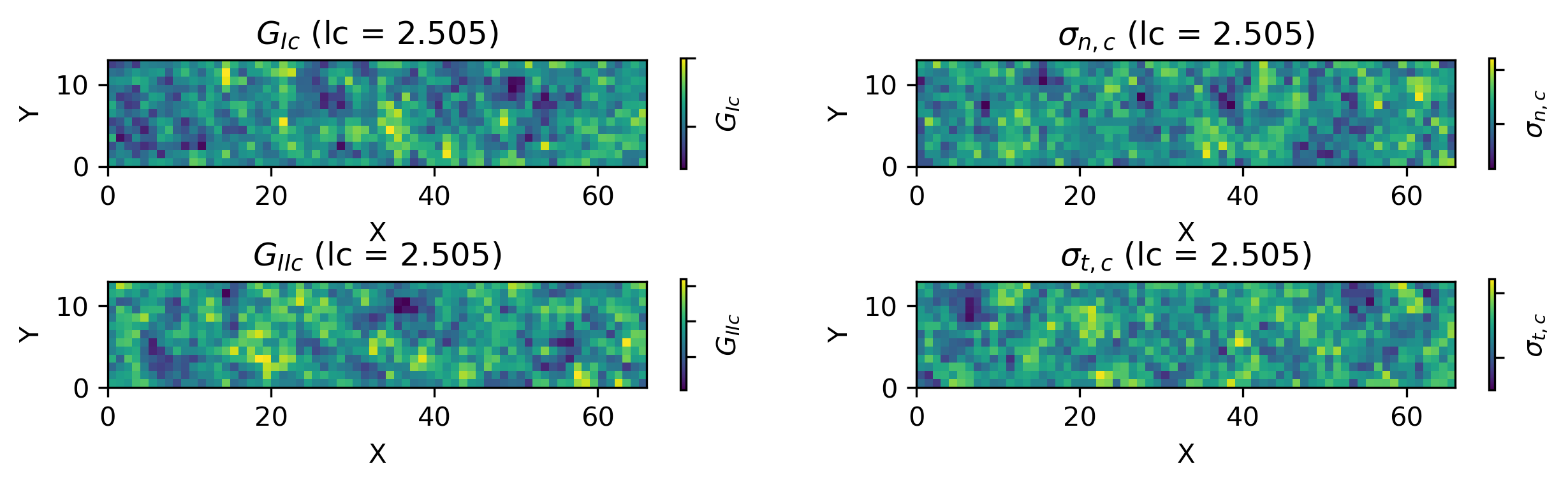}
        \caption{$\ell_{c3}$}
        \label{fig:EXP3}
    \end{subfigure}
    \caption{Samples of random fields generated using the \textbf{Exponential} correlation function for three correlation lengths.}
    \label{fig:EXP_3x1}
\end{figure}

\begin{figure}[H]
    \centering
    \begin{subfigure}[t]{0.9\textwidth}
        \centering
        \includegraphics[width=\linewidth]{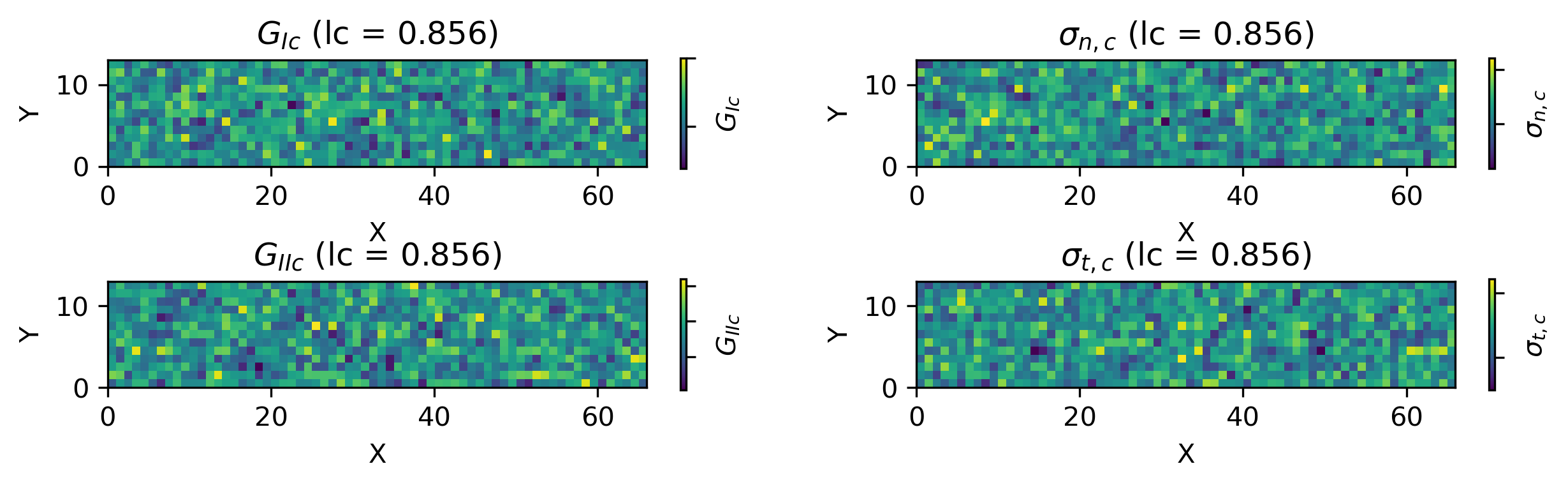}
        \caption{$\ell_{c1}$}
        \label{fig:Matern3_1}
    \end{subfigure}
    \vspace{0.1em}
    \begin{subfigure}[t]{0.9\textwidth}
        \centering
        \includegraphics[width=\linewidth]{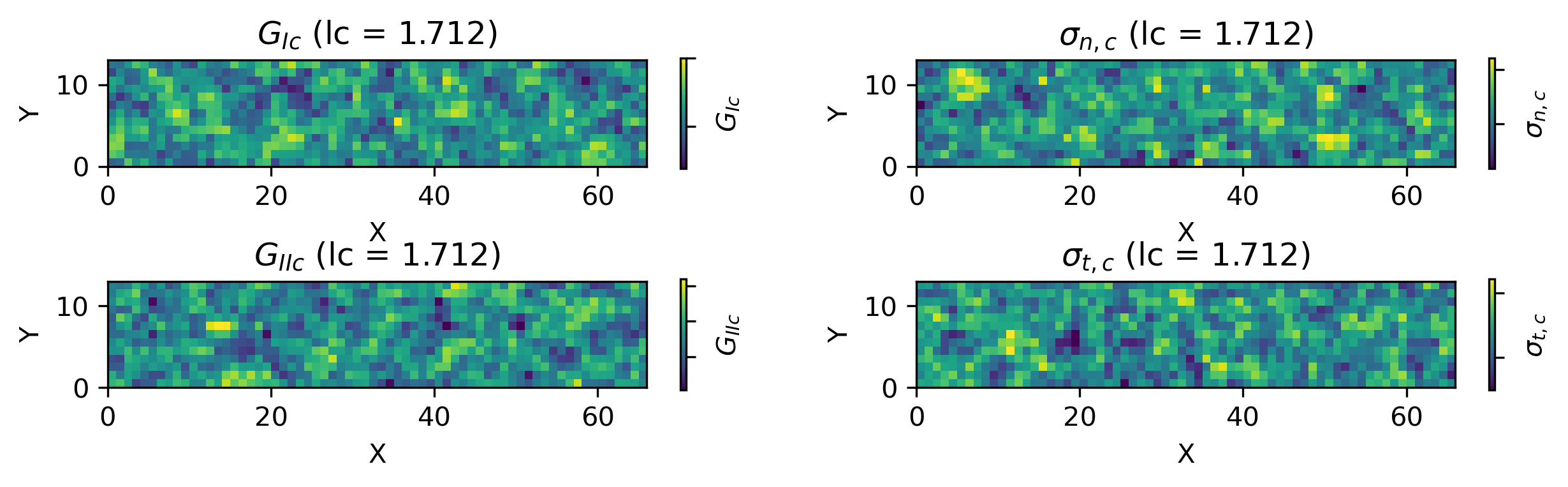}
        \caption{$\ell_{c2}$}
        \label{fig:Matern3_2}
    \end{subfigure}
    \vspace{0.1em}
    \begin{subfigure}[t]{0.9\textwidth}
        \centering
        \includegraphics[width=\linewidth]{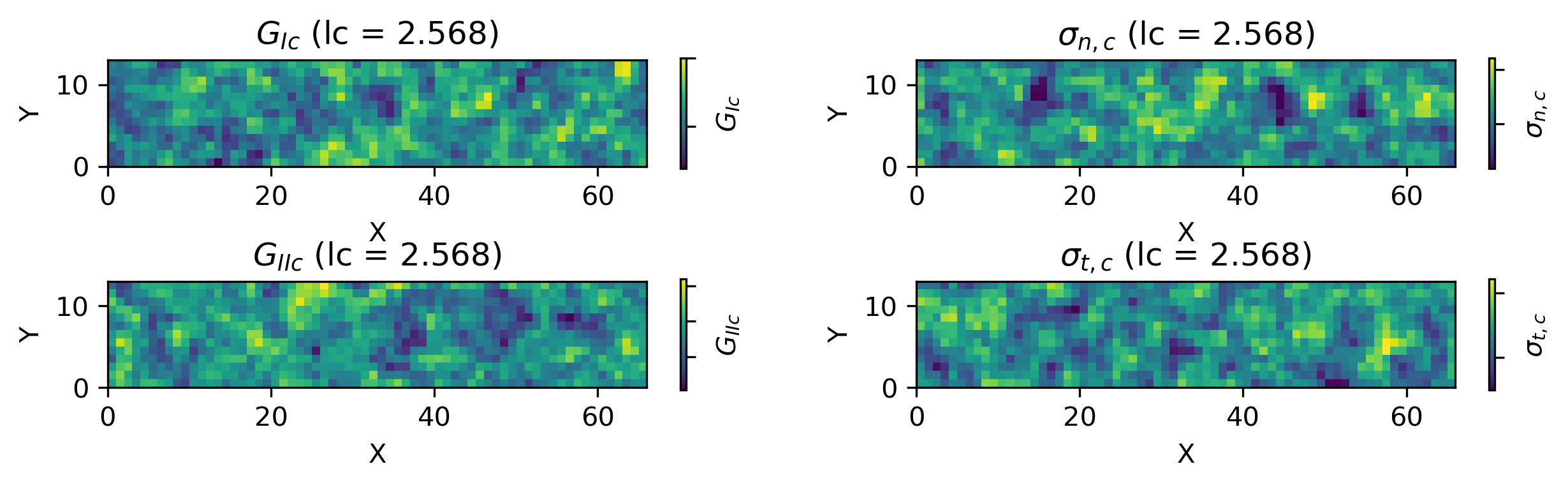}
        \caption{$\ell_{c3}$}
        \label{fig:Matern3_3}
    \end{subfigure}
    \caption{Samples of random fields generated using the \textbf{Matérn 3/2} correlation function for three correlation lengths.}
    \label{fig:Matern3_3x1}
\end{figure}

\begin{figure}[H]
    \centering
    \begin{subfigure}[t]{0.9\textwidth}
        \centering
        \includegraphics[width=\linewidth]{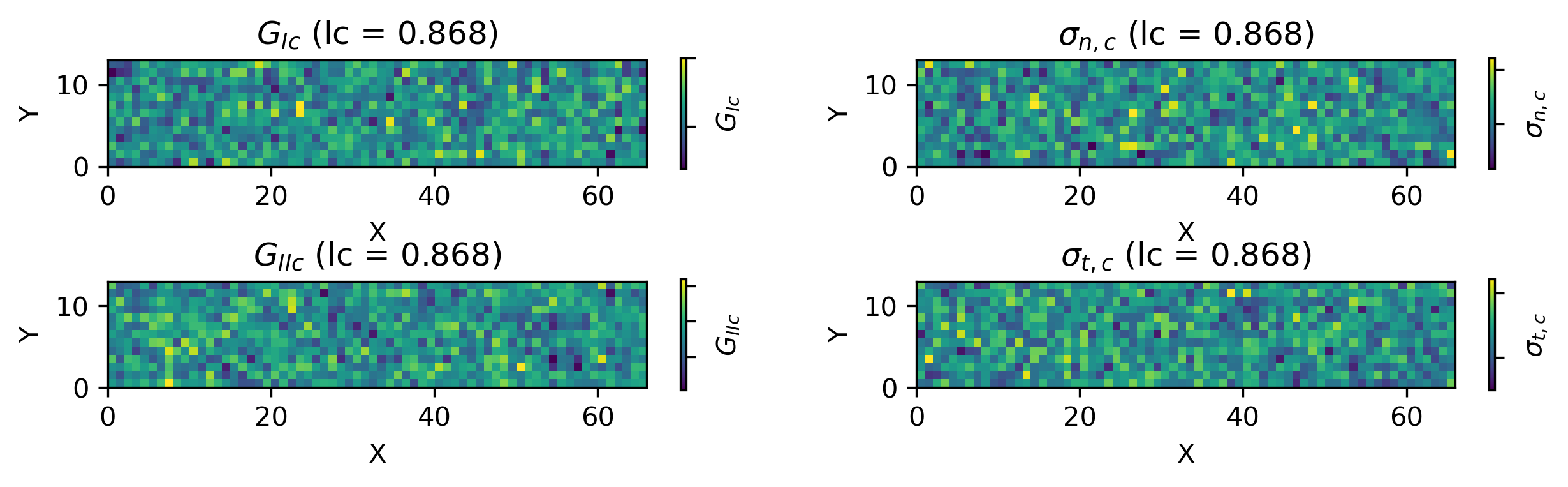}
        \caption{$\ell_{c1}$}
        \label{fig:Matern5_1}
    \end{subfigure}
    \vspace{0.1em}
    \begin{subfigure}[t]{0.9\textwidth}
        \centering
        \includegraphics[width=\linewidth]{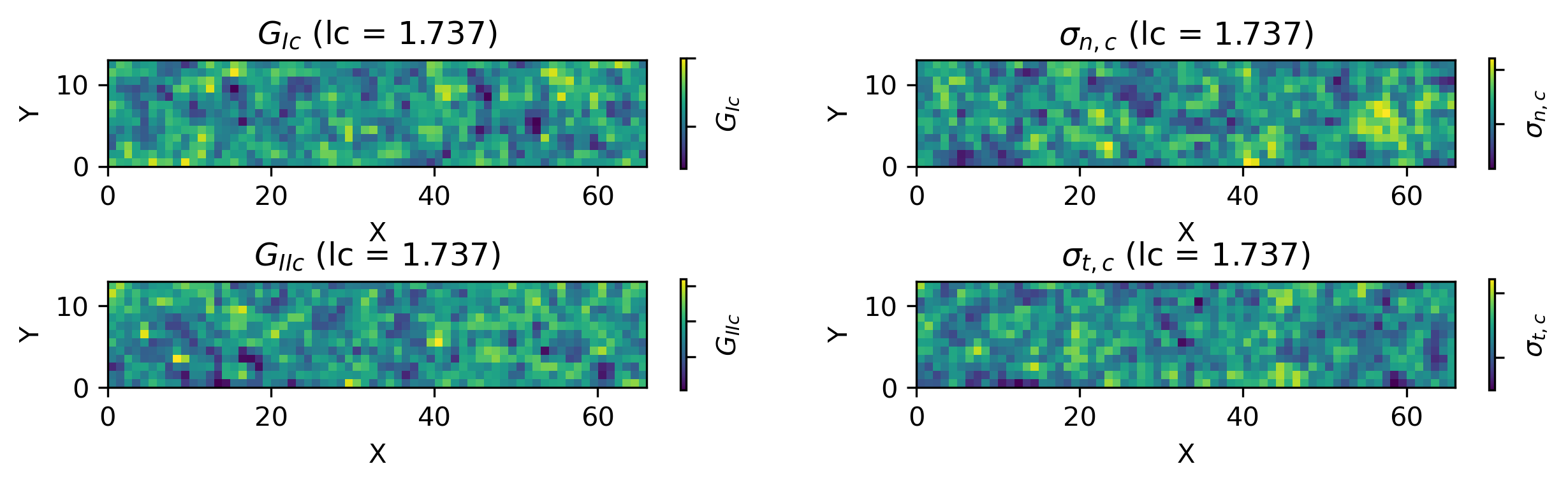}
        \caption{$\ell_{c2}$}
        \label{fig:Matern5_2}
    \end{subfigure}
    \vspace{0.1em}
    \begin{subfigure}[t]{0.9\textwidth}
        \centering
        \includegraphics[width=\linewidth]{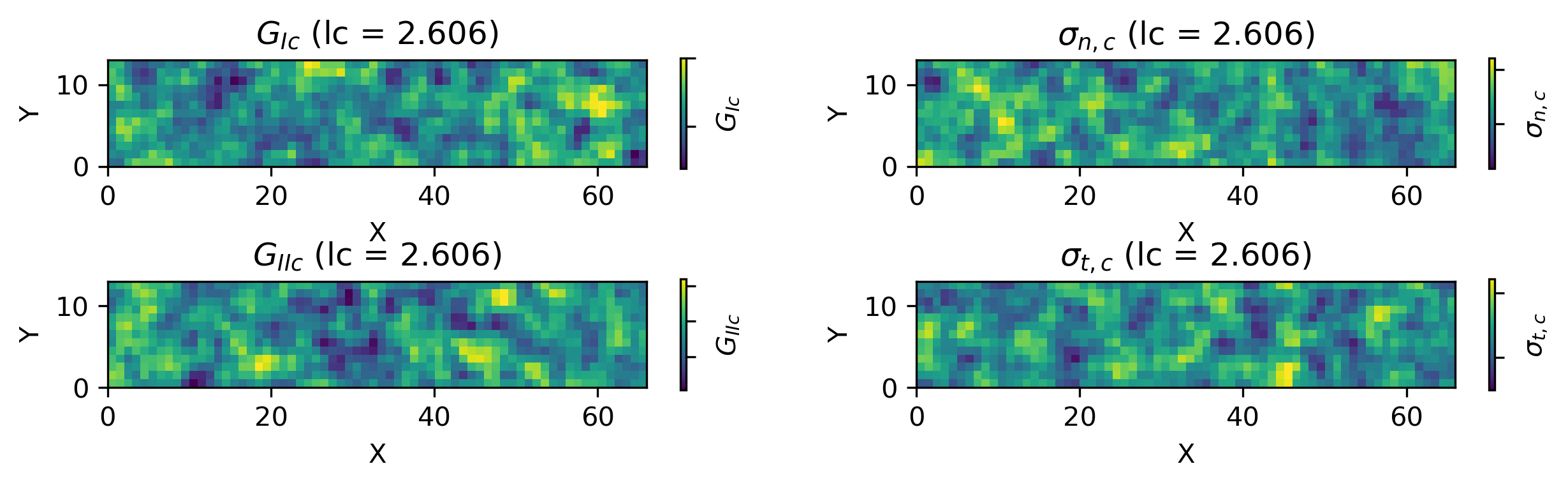}
        \caption{$\ell_{c3}$}
        \label{fig:Matern5_3}
    \end{subfigure}
    \caption{Samples of random fields generated using the \textbf{Matérn 5/2} correlation function for three correlation lengths.}
    \label{fig:Matern5_3x1}
\end{figure}

Figures ~\ref{fig:EXP_3x1} to \ref{fig:Matern5_3x1} show realizations of stationary Gaussian random fields generated with three different correlation functions and three correlation lengths $\ell_{c1}$, $\ell_{c2}$, and $\ell_{c3}$.

\section{Influence of correlation function regularity and correlation length on ENF simulations}
\label{ENFTest}

\begin{figure}[H]
    \centering
    \begin{subfigure}[b]{1\textwidth}
        \centering
        \hspace*{-2cm}
        \includegraphics[width=1.3\textwidth]{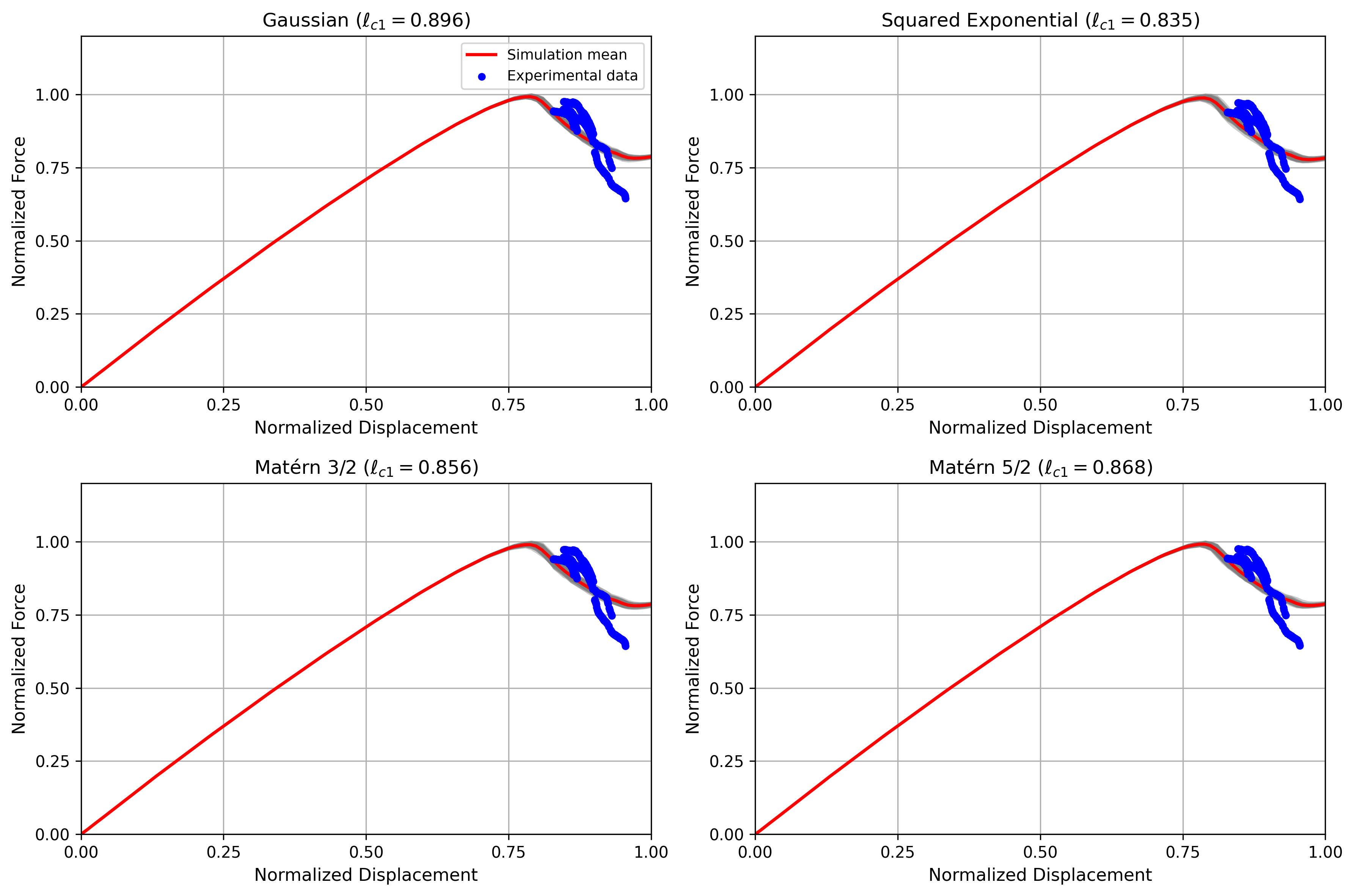}
        \caption{Correlation length $\ell_{c1}$.}
        \label{fig:enf1}
    \end{subfigure}
\end{figure}
\begin{figure}[H]
    \ContinuedFloat
    \centering
    \begin{subfigure}[b]{1\textwidth}
        \centering
        \hspace*{-2cm}
        \includegraphics[width=1.3\textwidth]{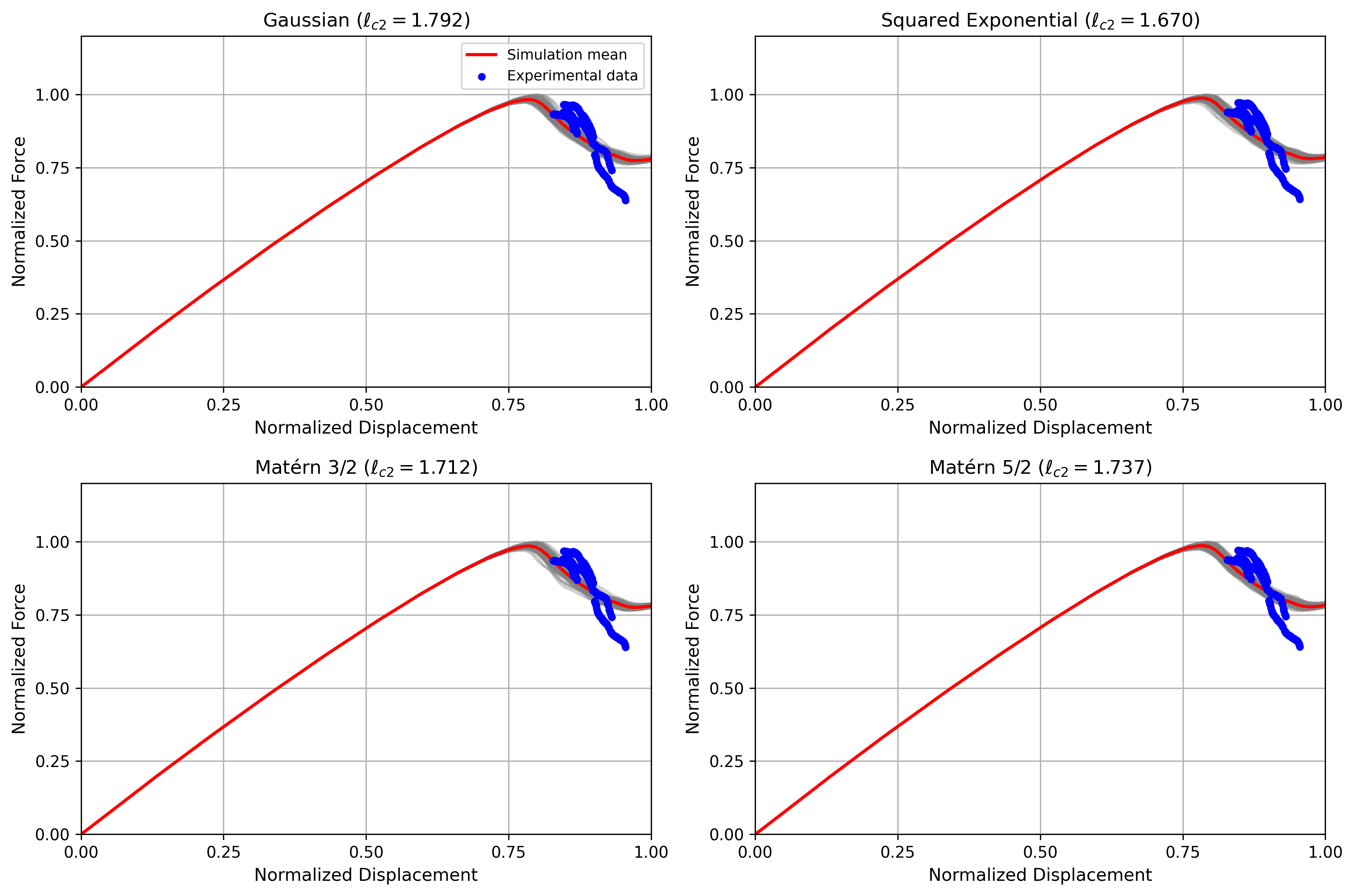}
        \caption{Correlation length $\ell_{c2}$.}
        \label{fig:enf2}
    \end{subfigure}
\end{figure}
\begin{figure}[H]
    \ContinuedFloat
    \centering
    \begin{subfigure}[b]{1\textwidth}
        \centering
        \hspace*{-2cm}
        \includegraphics[width=1.3\textwidth]{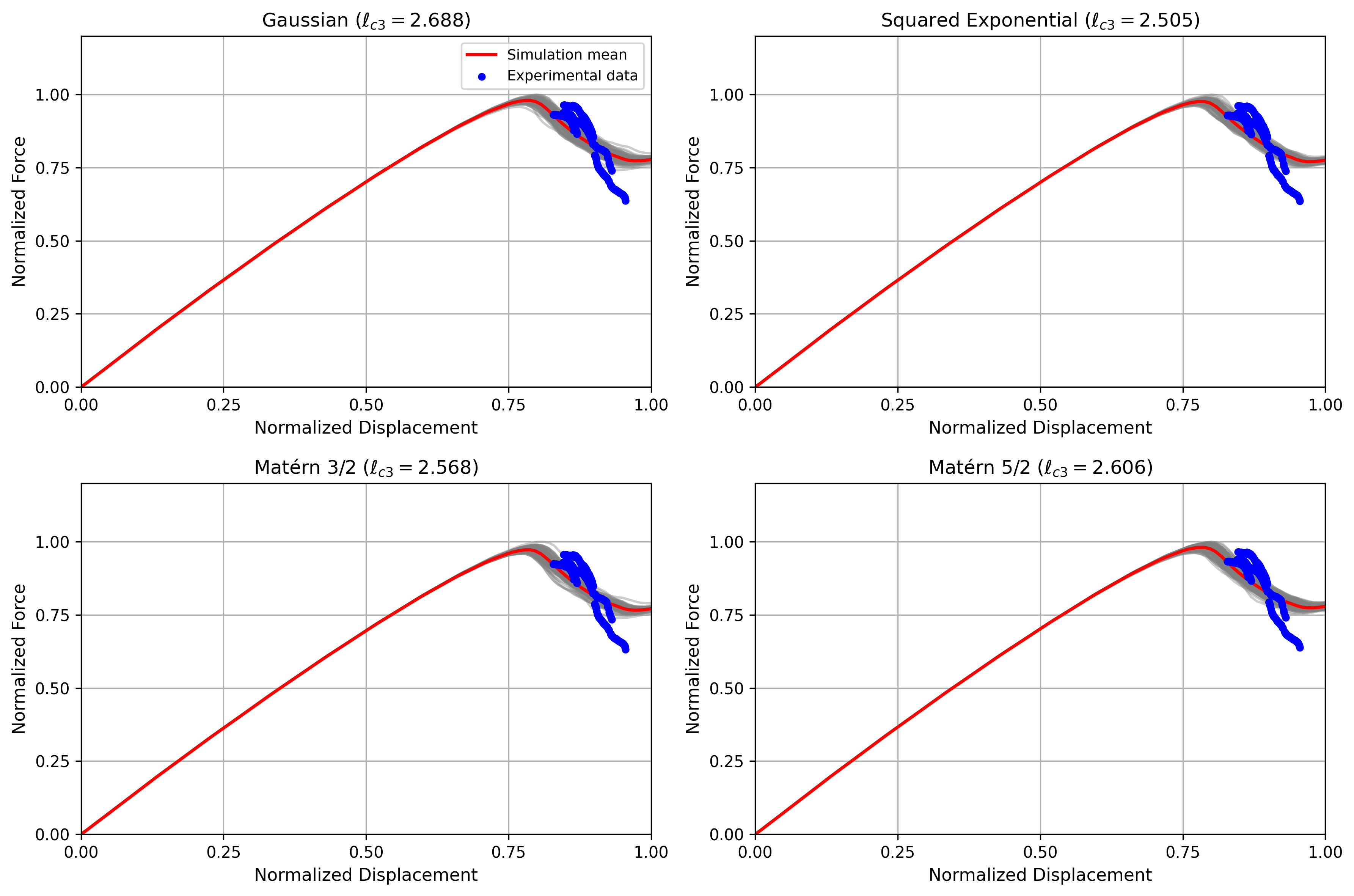}
        \caption{Correlation length $\ell_{c3}$.}
        \label{fig:enf3}
    \end{subfigure}

    \caption{Normalized force–displacement response of ENF simulations comparing different spatial correlation functions across three correlation lengths: (a) $\ell_{c1}$, (b) $\ell_{c2}$, and (c) $\ell_{c3}$. Experimental data are shown for reference.}
    \label{fig:enf_all}
\end{figure}




\end{appendices}


\bibliography{MaBib}

\end{document}